\documentclass{aastex631}

\usepackage{amsthm,amsmath,amssymb}
\usepackage{lipsum}
\usepackage{float}

\begin{document}

\title{Reanalyzing the Light Curves and Absolute Parameters of Twenty Contact Binary Stars\\
Using TESS Data}

\author{Ehsan Paki}
\affiliation{Binary Systems of South and North (BSN) Project, Tehran, Iran}

\author{Atila Poro}
\altaffiliation{atilaporo@bsnp.info}
\affiliation{Binary Systems of South and North (BSN) Project, Tehran, Iran}

%%%%%%%%%%%%%%%%%%%%%%%%%%%%%%%%%%%%%%%%%%%%%%%%%%%%%%
\begin{abstract}
Reanalyzing contact binaries with space-based photometric data and investigating possible parameter changes can yield accurate samples for theoretical studies. We investigated light curve solutions and fundamental parameters for twenty contact binary systems. The most recent Transiting Exoplanet Survey Satellite (TESS) data is used to analyze. The target systems in the investigation have an orbital period of less than 0.58 days. Light curve solutions were performed using the PHysics Of Eclipsing BinariEs (PHOEBE) Python code version 2.4.9. The results show that systems had various mass ratios from $q=0.149$ to $q=3.915$, fillout factors from $f=0.072$ to $f=0.566$, and inclinations from $i=52.8^{\circ}$ to $i=87.3^{\circ}$. The effective temperature of the stars was less than 7016 K, which was expected given the features of most contact binary stars. Twelve of the target systems' light curves were asymmetrical in the maxima, showing the O'Connell effect, and a starspot was required for light curve solutions. The estimation of the absolute parameters of the binary systems was presented using the $a-P$ empirical relationship and discussed. The orbital angular momentum ($J_0$) of the systems was calculated. The positions of the systems were also depicted on the $M-L$, $M-R$, $q-L_{ratio}$, $M_{tot}-J_0$, and $T-M$ diagrams.
\end{abstract}


%%%%%%%%%%%%%%%%%%%%%%%%%%%%%%%%%%%%%%%%%%%%%%%%%%%%
\section{Introduction}
Eclipsing binaries are important in investigating stellar formation and structure, examining stellar evolution theories, and determining stars' physical properties. Binary systems were first classified into three types based on the light curves' shapes: Algol (EA), $\beta$ Lyrae (EB), and W Ursae Majoris (W UMa, EW). Then, a more precise classification was provided by \cite{1959cbs..book.....K}[1], which was based on Roche geometry, binary systems were classified as detached, semi-detached, and contact. Contact binary systems are one of the most interesting kinds of stellar binaries among observers and researchers. Contact systems' stars have filled their Roche lobes (\citealt{1959cbs..book.....K}[1]), and the temperature difference between these stars is close to each other (\citealt{kuiper1941}[2]). Some of them are known as low-mass and Low-Temperature Contact Binaries (LTCBs) systems (\citealt{2005ApJ...629.1055Y}[3]).

According to the \cite{binnendijk1970}[4] study, contact binaries can be divided into A and W subtypes. The more massive component has a higher effective temperature in the A-subtype, and if the less massive component has a higher effective temperature, it is classified as a W-subtype. These subtypes are still being discussed, and for a better understanding, it is necessary to determine and analyze a large number of contact systems in terms of fundamental parameters.

The stars of contact systems are transferring mass to each other (\citealt{1979ApJ...231..502L}[5]), and in this process, their orbital periods can be changed. The orbital period of contact systems plays a role in relations with absolute parameters, and they are effective in the evolutionary process of these systems (\citealt{2021ApJS..254...10L}[6], \citealt{2022MNRAS.514.5528L}[7]). There have been several studies conducted on the upper and lower cut-offs of these systems' orbital periods (\citealt{2020MNRAS.497.3493Z}[8]). The investigations show that contact systems' orbital periods usually lie between 0.2 and 0.6 days.

One of the prominent features in many contact systems is the presence of starspot(s) induced by the stars’ magnetic activities. Hot or cold starspots, which are required for the light curve solution, are explained by the O'Connell effect (\citealt{1951PRCO....2...85O}[9]). The effect of starspots can be seen on the asymmetric maxima of the light curve.

There are unsolved issues related to contact binary systems, parameter relationships, and the evolution of stars (\citealt{2024RAA....24a5002P}[10]). This requires precisely defined elements from further contact binary samples (\citealt{2021AJ....162...13L}[11]). Even in the case of well-studied systems, it will be important to carry out investigations due to issues with phasing, challenging novel discoveries, and evolutionary status.

In the following sections, we present the general specifications of target systems in catalogs and literature (Section 2), light curve solutions of 20 contact binary systems (Section 3), estimating absolute parameters (Section 4), and finally a discussion and conclusion (Section 5).

%%%%%%%%%%%%%%%%%%%%%%%%%%%%%%%%%%%%%%%%%%%%%%%%%%
\vspace{1cm}
\section{Target Systems and Dataset}
\subsection{Systems' Selection}
We considered twenty contact binary systems for light curve analysis and absolute parameters' estimation. The selected studies for these target systems have considered the mass ratio based on spectroscopic results. We used these mass ratios as the initials for the light curve analysis. Due to the passage of time compared to some selected studies and spectroscopic quality, the final mass ratios may change slightly in this study.

Except for DY Cet (\citealt{2022RAA....22e5013Y}[12]), which has performed light curve analysis using TESS Sector 4 data, other systems have been studied previously using just ground-based data. Some parameters can be obtained from spectroscopic data analysis; however, most of the light curve elements are determined from photometric data. Therefore, the accuracy of photometric data has a significant impact on the results of the light curve solutions and then on the estimation of absolute parameters. Therefore, by using TESS data of suitable quality, some parameters may be obtained with more appropriate accuracy.

\subsection{TESS Observations}
TESS provides high accuracy and high time resolution light curves of contact binary stars that promote scientific studies. The main goal of the TESS mission is to detect and classify exoplanets, and each observation sector takes about 28 days. We used TESS data for light curve analysis in this study (\citealt{2010AAS...21545006R}[13], \citealt{2018AJ....156..102S}[14]). TESS data are available at the Mikulski Archive for Space Telescopes (MAST\footnote{\url{http://archive.stsci.edu/tess/all products.html}}). If each of the systems had several sectors in TESS, we selected the most recent available sector with good-quality data for the light curve analysis. The sector used for each system is listed in Table \ref{tab1}, and all of them were observed at a 120-second exposure time.

\begin{table*}
\caption{Coordinates and other characteristics of target binary systems.}
\centering
\begin{center}
\footnotesize
\begin{tabular}{c c c c c c c c c}
 \hline
System	&	RA.(J2000)	&	DEC.(J2000)	&	$d$(pc)	&	$V$(mag)	& $A_V$ &	$t_0(BJD_{TDB})$	&	P(day)	&	TESS Sector\\
\hline
AC Boo	&	14 56 28.3364	&	+46 21 44.0691	&	155.42(36)	&	10.27(21)	&	0.029(1)	&	2452499.9507	&	0.35245	&	50	\\
AQ Psc	&	01 21 03.5557	&	+07 36 21.6178	&	133.27(37)	&	8.55(18)	&	0.079(1)	&	2453653.7169	&	0.47560	&	43	\\
BI CVn	&	13 03 16.4093	&	+36 37 00.6406	&	221.26(2.11)	&	10.41(22)	&	0.031(1)	&	2444365.2503	&	0.38416	&	49	\\
BX Dra	&	16 06 17.3670	&	+62 45 46.0898	&	520.27(4.65)	&	10.68(22)	&	0.048(1)	&	2449810.5906	&	0.57902	&	58	\\
BX Peg	&	21 38 49.3911	&	+26 41 34.2134	&	149.34(56)	&	10.88(23)	&	0.076(1)	&	2455873.3966	&	0.28042	&	55	\\
CC Com	&	12 12 06.0379	&	+22 31 58.6828	&	71.38(11)	&	11.21(23)	&	0.035(1)	&	2453012.8637	&	0.22069	&	49	\\
DY Cet	&	02 38 33.1803	&	-14 17 56.7219	&	186.88(49)	&	9.47(20)	&	0.046(1)	&	2453644.7385	&	0.44079	&	4	\\
EF Boo	&	14 32 30.5386	&	+50 49 40.6868	&	160.93(34)	&	9.63(20)	&	0.021(1)	&	2452500.2238	&	0.42052	&	50	\\
EX Leo	&	10 45 06.7720	&	+16 20 15.6771	&	97.10(24)	&	8.91(19)	&	0.037(1)	&	2448500.0087	&	0.40860	&	46	\\
FU Dra	&	15 34 45.2133	&	+62 16 44.3332	&	159.56(50)	&	10.68(22)	&	0.030(1)	&	2448500.2637	&	0.30672	&	51	\\
HV Aqr	&	21 21 24.8100	&	-03 09 36.8855	&	154.24(2.32)	&	9.85(21)	&	0.090(1)	&	2452500.2191	&	0.37446	&	55	\\
LO And	&	23 27 06.6850	&	+45 33 22.0263	&	290.62(3.82)	&	11.26(23)	&	0.213(2)	&	2456226.6800	&	0.38044	&	57	\\
OU Ser	&	15 22 43.4748	&	+16 15 40.7337	&	53.29(7)	&	8.14(17)	&	0.024(1)	&	2448500.2787	&	0.29677	&	51	\\
RW Com	&	12 33 00.2840	&	+26 42 58.3618	&	107.60(25)	&	11.05(23)	&	0.026(1)	&	2454918.7048	&	0.23735	&	49	\\
RW Dor	&	05 18 32.5451	&	-68 13 32.7780	&	123.57(17)	&	11.00(23)	&	0.087(1)	&	2453466.5302	&	0.28546	&	67	\\
RZ Com	&	12 35 05.0595	&	+23 20 14.0278	&	208.26(1.70)	&	10.34(22)	&	0.032(1)	&	2458253.6304	&	0.33851	&	49	\\
TW Cet	&	01 48 54.1435	&	-20 53 34.5917	&	152.71(46)	&	10.40(22)	&	0.022(1)	&	2454476.6173	&	0.31685	&	3	\\
UV Lyn	&	09 03 24.1259	&	+38 05 54.5972	&	143.65(36)	&	9.60(20)	&	0.036(1)	&	2453407.3606	&	0.41498	&	21	\\
UX Eri	&	03 09 52.7437	&	-06 53 33.5110	&	237.40(1.00)	&	11.15(23)	&	0.142(1)	&	2454828.6698	&	0.44529	&	4	\\
VW Boo	&	14 17 26.0325	&	+12 34 03.4469	&	150.05(38)	&	10.49(22)	&	0.037(1)	&	2452840.6121	&	0.34232	&	50	\\
\hline
\end{tabular}
\end{center}
\label{tab1}
\end{table*}

\subsection{General Features}
The selected contact binary systems have orbital periods ranging from 0.22 to 0.58 days, their apparent magnitude is $8.14^{mag}$ to $11.26^{mag}$, and the effective temperature of star 1 ($T_1$) is 4300 to 6980 K in the reference studies.

Table \ref{tab1} contains the names of the selected systems along with their general characteristics. Therefore, in Table \ref{tab1}, the RA. and DEC. of the systems from the Simbad\footnote{\url{https://simbad.u-strasbg.fr/simbad/}} database, the distance obtained from the Gaia DR3\footnote{\url{https://gea.esac.esa.int/archive/}} parallax, the apparent magnitude of the system from the All Sky Automated Survey (ASAS) catalog, the time of minima ($BJD_{TDB}$) and the orbital period from the Variable Star Index (VSX)\footnote{\url{https://www.aavso.org/vsx/index.php?view=search.top}} database, and the last column, the used TESS sector, were given. We used \cite{2011ApJ...737..103S}[15] study, and the DUST-MAPS Python package developed by \cite{2019ApJ...887...93G}[16] to determine the extinction coefficient ($A_V$) along with its uncertainty.

\subsection{Introducing the Systems}
$\bullet$ AC Boo: \cite{herz1995marcus}[17] discovered the variability of AC Boo. The system was first classified as a W UMa-type by \cite{zessewitsch1956}[18], who also tried to determine the system's orbital period. The rectification method was employed by \cite{1964ZA.....60..222M}[19] to analyze the light curves. \cite{1974IBVS..904....1M}[20], \cite{1977AAS...29...57M}[21], and \cite{1978AA....63..193M}[22] classified the system as a W-subtype with a $q=3.57$, $i=85.47^{\circ}$, and $T_1=6100$ K. \cite{1983AAS...52..463S}[23] investigated and suggested two values, $q=0.31$ and $q=0.28$, for the mass ratio. \cite{2010IBVS.5951....1N}[24] studied this system, presenting a new ephemeris in addition to a light curve analysis.

$\bullet$ AQ Psc: \cite{1982IBVS.2073....1S}[25] discovered AQ Psc as a short orbital period W UMa-type binary system with the F8V spectral type. \cite{1999AJ....118..515L}[26] investigated the radial velocities of AQ Psc and determined the mass ratio to be $q=0.226$. \cite{2005ApSS.296..277Y}[27] conducted the first detailed photometric light curve study on AQ Psc and determined the system's parameters. According to other investigations like \cite{2006AJ....131.2986P}[28], AQ Psc is a system where the third component is probable. \cite{2006PASA...23..154D}[29] derived the physical parameters of this system and concluded it is an A-type overcontact configuration system. \cite{2011MNRAS.412.1787D}[30] studied this system using the ASAS-3 project data.

$\bullet$ BI CVn: \cite{kippenhahn1955}[31] discovered the variability of BI CVn. Then, \cite{1960ATsir.215...20F}[32] raised doubts about W UMa-type variability. This contact binary system was classified as a W-subtype by \cite{1987ApSS.135..169D}[33], who refunded the orbital period to 0.3841692 days. By contrast, the \cite{1988AcApS...8..259L}[34] and \cite{1996AA...311..523M}[35] studies reported the mass ratio of $q=0.87$ and $q_{sp}=0.41$, respectively, and assigned the categorization to A-subtype. Multiband light curves were acquired by \cite{2008AJ....136.2493Q}[36], who also provided a photometric solution. \cite{2008AJ....136.2493Q}[36] concluded that the spectroscopic mass ratio reported by \cite{1988AcApS...8..259L}[34] supported the A-type classification, while the temperature differential between the components suggested the nature of a W-type. \cite{2014NewA...29...57N}[37] presented photometric and spectroscopic observations and concluded this system is a W-type overcontact binary.

$\bullet$ BX Dra: This system was discovered as a short-period variable by \cite{strohmeier1958}[38]. \cite{1990PASP..102..124S}[39] suggested that BX Dra be an ellipsoidal-type variable. \cite{2004AJ....127.1712P}[40] obtained double-line radial-velocity curves of BX Dra and found that this system could be an A-subtype contact binary. \cite{2007ApSS.312..151S}[41], \cite{2009IBVS.5872....1K}[42], and \cite{2010MNRAS.408..464Z}[43] observed and analyzed this system, and they considered a third light in their solutions. \cite{2013PASJ...65....1P}[44] performed long-term photometric observations for this system in several filters. They suggested that BX Dra is probably a triple system.

$\bullet$ BX Peg: \cite{shapley1934variable}[45] classified BX Peg as a variable. Based on a radial velocity investigation by \cite{1991AJ....102.1171S}[46], the spectroscopic mass ratio value $q_{sp}=0.372$ was determined. Additionally, \cite{2009PASP..121.1366L}[47] reported an orbital period decrease and a possible third body in this system. \cite{2013JAVSO..41..227A}[48] collected photometric data in three filters and analyzed light curves. \cite{2013JAVSO..41..227A}[48] also studied the orbital period variations of BX Peg. \cite{2015NewA...41...17L}[49] observed and presented multiband light curves, and concluded that BX Peg is a W-type shallow contact binary system.

$\bullet$ CC Com: This system was discovered by \cite{1964AN....288...49H}[50]. CC Com was investigated with photometric observations by \cite{1976PASP...88..777R}[51], \cite{1982ApSS..86..107B}[52], \cite{1985ApJS...58..413B}[53], \cite{1988ApSS.141..199Z}[54], \cite{1989ApJ...343..909L}[55], and \cite{2010MNRAS.408..464Z}[43]. Furthermore, there are some spectroscopic studies. So, the spectroscopic mass ratio was reported as $q_{sp}=0.52(3)$, $q_{sp}=0.47(4)$, and $q_{sp}=0.53(1)$ in the \cite{1977PASP...89..684R}[56], \cite{1983MNRAS.203....1M}[57], and \cite{2007AJ....133.1977P}[58] studies, respectively. Furthermore, the \cite{2011AN....332..626K}[59] study obtained the parameters by analyzing their photometric observations with archival spectroscopic data.

$\bullet$ DY Cet: DY Cet was identified as an eclipsing binary system and included as a W UMa type system in the Hipparcos and Tycho catalogs (\citealt{1997AA...323L..49P}[60]). \cite{2004AA...416.1097S}[61] determined some light elements of DY Cet using the Hipparcos light curve as $f=0.2$, $i=77.5^{\circ}$, and $q=0.45$. \cite{2009AJ....137.3655P}[62] analyzed the radial velocity of DY Cet, estimated the mass ratio as $q=0.356(9)$, and concluded that this system is an A-subtype W UMa. Using ASAS data, \cite{2011MNRAS.412.1787D}[30] conducted a light curve study of the system and derived the fundamental astrophysical properties of the component stars. \cite{2022RAA....22e5013Y}[12] used TESS observations Sector 4 to analyze the light curve together with the published radial velocity data.

$\bullet$ EF Boo: This system was discovered by the Hipparcos mission. A photoelectric light curve analysis by \cite{1999IBVS.4811....1S}[63] in three filters, estimated the mass ratio of EF Boo to be $q=1.75$. The first spectroscopic observations of EF Boo were made by \cite{2001AJ....122.1974R}[64] and the mass ratio reported was $q=0.512(8)$, and it was considered a W-type system. \cite{2004AA...416.1097S}[61] studied EF Boo and obtained a mass ratio of $q=0.45$, fillout factor of $f=0.2$, and inclination of $i=77.5^{\circ}$. EF Boo has also been investigated in several studies about the O'Connell effect and the requirement for any starspot in light curve solutions (e.g., \citealt{2000AAS...197.4803G}[65], \citealt{2001IBVS.5033....1O}[66]).
\cite{2005AcA....55..123G}[67] studied EF Boo. They reported a fillout factor of $f=0.18$ and a mass ratio of $q=1.871$. \cite{2005AcA....55..123G}[67] concluded EF Boo is a W-type contact binary system. \cite{2017NewA...55...13Y}[68] published another study on this system, focusing on the orbital period variations.

$\bullet$ EX Leo: This system was discovered by the Hipparcos mission (\citealt{perryman1997hipparcos}[69]). According to spectroscopic analysis, \cite{2001AJ....122..402L}[70] determined that the system's mass ratio is $q=0.199(36)$ indicating that EX Leo's stars belong to the F6V spectral category. The light curve solution by the \cite{2002IBVS.5258....1P}[71] study obtained an orbital inclination of $i=61^{\circ}$ and a fillout factor of $f=0.31$. There have been some unsuccessful attempts to find a possible third body in this system (\citealt{2006AJ....132..650D}[72], \cite{2006AJ....131.2986P}[28]). The multicolor light curve modeling findings for EX Leo were provided by the \cite{2010MNRAS.408..464Z}[43]. They used a combination of photometric and radial velocity for solutions, and the absolute parameters of the components have been determined.

$\bullet$ FU Dra: The variability of FU Dra was first reported by \cite{1922AN....216..159B}[73]. This system was discovered on the Hipparcos mission. \cite{2000AJ....120.1133R}[74] reported that FU Dra is a W-type contact binary system with spectral type F8V. The O'Connell effect in this system has been the subject of many investigations (e.g. \citealt{2005AcA....55..389Z}[75], \citealt{2006JAVSO..34..165K}[76]). \cite{2012PASJ...64...48L}[77] analyzed the light curve of the system using multicolor filters. They concluded FU Dra is a W-type contact binary and has a high fill-out factor ($f=0.267$).

$\bullet$ HV Aqr: The light variability of HV Aqr was discovered by \cite{1992IBVS.3723....1H}[78]. \cite{1992IBVS.3785....1S}[79] observed and reported a G5 spectral type for HV Aqr. The first light curve analysis of HV Aqr was done by \cite{1992IBVS.3798....1R}[80], who proposed that this system has an extremely low mass ratio ($q=0.146$) with an inclination of $i=78.3^{\circ}$ and a fillout factor of $f=0.475$. \cite{2000IBVS.4951....1M}[81] and \cite{2000AJ....120.1133R}[74] analyzed the light curves and reported $q=0.18$ and $q=0.145(50)$, respectively. \cite{2007ASPC..370..279G}[82] obtained the mass ratio $q=0.151$ and they also found the presence of the O'Connell effect in the light curves. \cite{2013NewA...21...46L}[83] presented photometric solutions using three filter observations. They reported that their determined light curves do not show the O’Connell effect. \cite{2013NewA...21...46L}[83] solutions suggested that
HV Aqr is a low mass ratio and deep contact binary system.

$\bullet$ LO And: \cite{1963IBVS...21....1W}[84] reported the discovery of LO And's variability. The first complete light curve was provided by \cite{1980IBVS.1767....1D}[85], who reported that LO And displays a typical EW-type luminosity variation. \cite{2005AN....326...43G}[86] derived the photometric solution and concluded LO And is an A-type contact binary with $f=0.306$ and $q=0.371$. \cite{2015IBVS.6134....1N}[87] performed both photometric and spectroscopic observations and reported the O’Connell effect. \cite{2015IBVS.6134....1N}[87] argued that LO And is a W-subtype contact binary. \cite{2021RAA....21..120H}[88] presented the geometric, photometric, and absolute parameters of LO And. They concluded that this is an A-type contact binary system with a cool third component.

$\bullet$ OU Ser: This system was discovered by Hipparcos. \cite{2000AJ....120.1133R}[74] used spectroscopic observations to investigate and result in reliable parameters. \cite{2000AJ....120.1133R}[74] classified this system as an A-type W UMa binary system. The light curve analysis was presented using ground-based data and in $UBV$ filters by \cite{2002CoSka..32...79P}[89] and concluded this system is a W-subtype. \cite{2011MNRAS.412.1787D}[30] presented light curve solutions and absolute parameters for OU Ser.

$\bullet$ RW Com: The variability of RW Com was first recognized by \cite{1923AJ.....35...44J}[90]. The system's first spectral observations were carried out by \cite{1950ApJ...111Q.658S}[91]. \cite{1980ApJS...43..339M}[92] used light curves that were observed up to the time of their study. \cite{1985AJ.....90..109M}[93] identified this system as a W-type contact binary based on their parameters and mass ratio estimation of $q=0.34$. \cite{1987ApSS.139..373S}[94] and \cite{2002AA...384..908Q}[95] studied the orbital period variations of RW Com and suggested the existence of a third body in the system. However, \cite{2006AJ....131.2986P}[28] and \cite{2007AJ....134.2353R}[96] did not detect a third body in their studies. Furthermore, broadening functions developed by \cite{2009AJ....137.3646P}[97] for radial velocity analysis showed no evidence of a third body in the system. \cite{2011AA...525A..66D}[98] analyzed new multicolor light curves and found $i=74.9^{\circ}$, $f=0.06$, and $q=0.47$1. \cite{2011MNRAS.412.1787D}[30] studied 62 binary systems including RW Com in $V$ filter.

$\bullet$ RW Dor: \cite{1908AnHar..60...87L}[99] discovered RW Dor as a variable star, which was then classified as a W UMa-type eclipsing binary by \cite{1928BAN.....4..153H}[100]. \cite{1989MNRAS.240..931M}[101] and \cite{1989AcA....39...27K}[102] investigated the light curves and separately determined the photometric elements. According to those solutions, RW Dor belongs to a W-type contact binary. \cite{1992MNRAS.255..285H}[103] presented the first spectroscopic observations of RW Dor. They found that RW Dor consisted of two stars of the K1 spectral type and determined a mass ratio of $q_{sp}=0.68$. \cite{2007AJ....133..169D}[104] presented another radial-velocity curve and obtained a spectroscopic mass ratio of $q_{sp}=0.63$. \cite{2019PASJ...71...34S}[105] analyzed the light curves and found RW Dor is a W-type shallow contact binary system. They reported a mass ratio of $q=1.587$ and based on the results, the hotter component is the less massive one.

$\bullet$ RZ Com: \cite{1932VeBB....9D...1G}[106] and \cite{payne1938variable}[107] reported an inclination of $i=81^{\circ}$ and classified RZ Com as a W UMa-type binary system. Based on the spectroscopic observations, \cite{1948ApJ...108..497S}[108] determined a mass ratio of $q=2.1$. \cite{1955AnAp...18..379K}[109] reported RZ Com's component masses of $0.8M_{\odot}$ and $1.6M_{\odot}$, spectral types of G9 and K0, respectively. \cite{2004NewA....9..273X}[110] studied RZ Com and obtained a mass ratio of $q=0.8$ and $q=2.2$ using the $q$-search method, but they could not determine which mass ratio was more accurate. \cite{2019IBVS.6266....1N}[111] presents photometric light curve and radial velocity observations. They concluded that RZ Com is a W-type overcontact binary system.

$\bullet$ TW Cet: \cite{1933AN....247..281H}[112] classified TW Cet as a variable star. TW Cet was identified by \cite{zessewitsch1944}[113] as an eclipsing binary of the W UMa-type after analyzing its light curve. \cite{1950ApJ...111R.658S}[114] conducted spectroscopic observations of TW Cet. \cite{1951AJ.....56...35C}[115] produced the first photoelectric observations of this binary system, obtaining complete light curves and finding that the secondary minimum is deeper than the primary one. Light curve analysis was done by \cite{1959ApJ...130..774A}[116] to determine photometric parameters. \cite{1982AAS...47..211R}[117] observed this system and concluded that it is a W-type with $i=83.27(49)$, $f=0.044$, and $q=1.664(30)$. TW Cet was one of the binary systems that \cite{2011MNRAS.412.1787D}[30] studied.

$\bullet$ UV Lyn: This system was discovered by Kippenhahn as a variable star (\citealt{strohmeier1960veroff}[118]). UV Lyn was classified by \cite{1973AAS...10..217B}[119] to be a W UMa-type binary with an asymmetry in the maxima and a period of 0.415 days. \cite{1982PASP...94..350M}[120] reported that UV Lyn is a contact binary system with an inclination of $i=68^{\circ}$ and a mass ratio of $q=0.526$. \cite{1995IBVS.4240....1Z}[121] and \cite{2001CoSka..31..129V}[122] were the two studies that found an increase in the orbital period of UV Lyn. They proposed mass transfer from the secondary to the primary component to explain the orbital period change. The first spectroscopic orbit of UV Lyn was determined by \cite{1999AJ....118..515L}[26] and reported $q=0.367$. \cite{2005AcA....55..389Z}[75] investigated UV Lyn using combined spectroscopic and multicolor photometric observations. They determined the fillout factor to be $f=0.18$ and a starspot based on the asymmetry in the maxima of the light curve.

$\bullet$ UX Eri: The light variability of the UX Eri was discovered by \cite{soloviev1937tadjik}[123]. Complete photoelectric light curves in two passbands were acquired and analyzed by \cite{1967AJ.....72...82B}[124]. \cite{1972AA....17....1M}[125] reanalyzed the light curves of \cite{1967AJ.....72...82B}[124] and determined $q=0.42$ as the photometric mass ratio. \cite{2000AJ....120.1133R}[74] provided the first radial velocity curves of UX Eri and determined the spectroscopic features of this binary system. \cite{2005AcA....55..123G}[67] analyzed the light curve and concluded that UX Eri is an overcontact binary with a small fillout factor ($f=0.13$). \cite{2006AJ....131.2986P}[28] found that UX Eri was a triple system with a faint tertiary component. \cite{2007AJ....134.1769Q}[126] reported solutions that confirmed UX Eri is a marginal W-type overcontact binary system with a low degree of overcontact. \cite{2011MNRAS.412.1787D}[30] presented a light curve analysis of UX Eri.

$\bullet$ VW Boo: \cite{1935AN....255..401H}[127] discovered this system. \cite{zessewitsch1944}[113] and \cite{zessewitsch1956}[18] observed VW Boo and reported it as a W UMa-type binary system. \cite{1990MNRAS.246...47R}[128] and \cite{2009AJ....137.3646P}[97] measured its radial velocities. \cite{1990MNRAS.246...47R}[128] concluded that VW Boo is in the transition phase from B-type to W UMa-type. According to the \cite{2002ApJ...568.1004Q}[129] investigation of this system, VW Boo was at the beginning of the overcontact phase. \cite{2011AJ....141..147L}[130] presented a light curve solution using multicolor photometric observations and confirmed that VW Boo is a shallow W-type contact binary system.

%%%%%%%%%%%%%%%%%%%%%%%%%%%%%%%%%%%%%%%%%%%%%%%%%%
\vspace{1cm}
\section{Light Curve Analysis}
The PHOEBE Python code version 2.4.9 (\citealt{2016ApJS..227...29P}[131], \citealt{2020ApJS..250...34C}[132]) and TESS filter were used to perform light curve analysis on 20 binary systems. Based on the appearance of the light curve, the short orbital period, and the results of previous studies, we selected contact mode for the systems' light curve solutions.

Assumed values for the bolometric albedo and gravity-darkening coefficients were $A_1=A_2=0.5$ (\citealt{1969AcA....19..245R}[133]) and $g_1=g_2=0.32$ (\citealt{1967ZA.....65...89L}[134]).
We used the \cite{castelli2004new}[135] method to model the stellar atmosphere, and the limb darkening coefficients were from the PHOEBE tables. The reflection effect in contact binary systems refers to the component's irradiation of each other. We considered this effect in the light curve analysis.

We set initial effective temperatures for stars and mass ratio from reference studies and analyzed the light curves of 20 target systems (Table \ref{tab2}). Parameter input values were taken from studies including spectroscopic data for these target systems. The optimization tool in PHOEBE was then utilized to improve the output of light curve solutions and obtain the final results. The five main parameters ($T_1$, $T_2$, $q$, $f$, $l_1$) were then processed for the optimization.

The well-known O'Connell effect appears by the asymmetry in the brightness of maxima in the light curve of eclipsing binary stars (\citealt{1951PRCO....2...85O}[9]). The most probable reason for this phenomenon is the existence of starspot(s) caused by the components' magnetic activity (\citealt{2017AJ....153..231S}[136]). Eight systems had symmetrical light curves and 12 systems needed a cold starspot for light curve analysis.

\begin{figure*}
   \centering
   \includegraphics[scale=0.134]{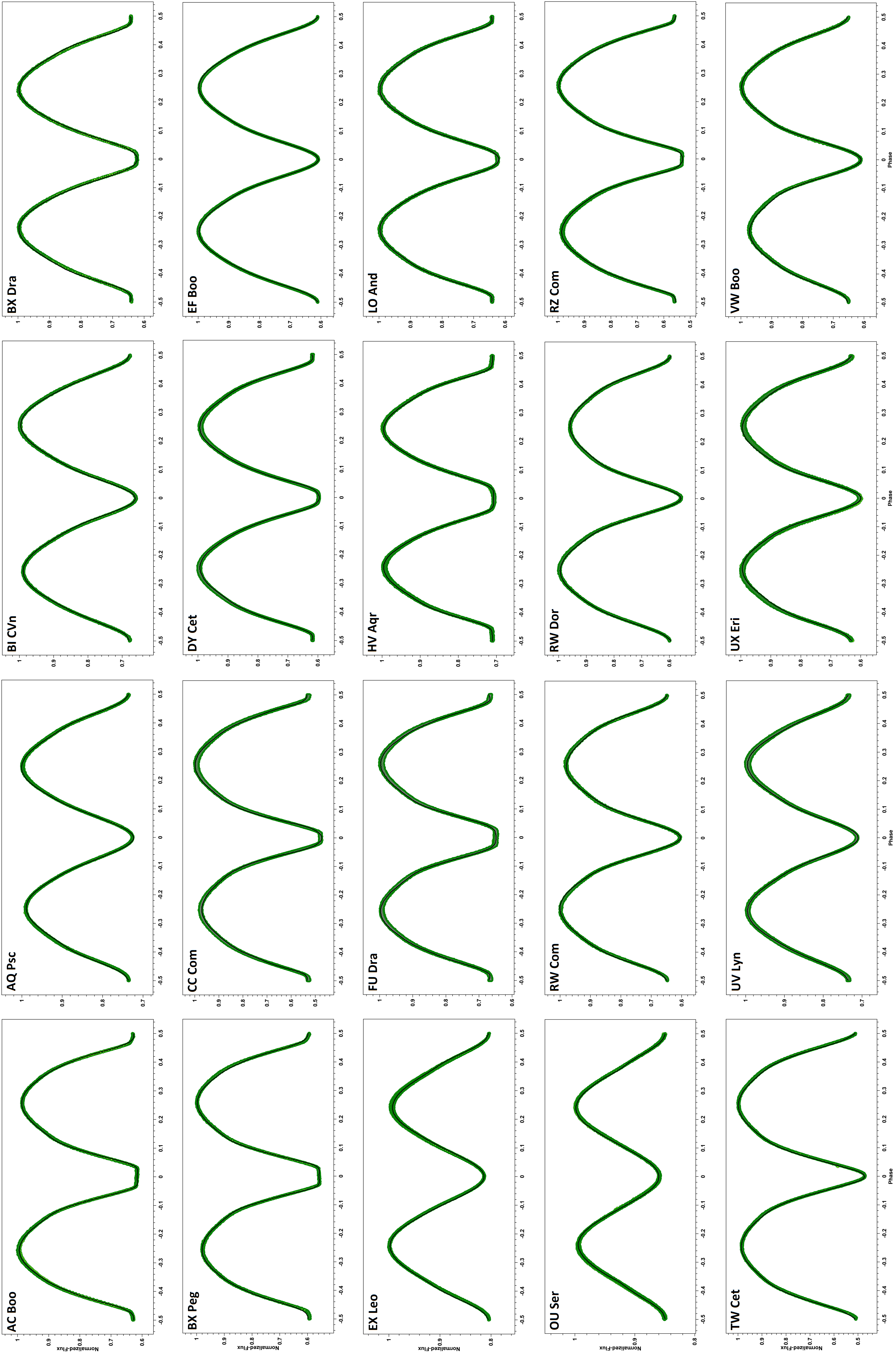}
    \caption{TESS data and synthetic light curves for the target binary systems.}
    \label{Fig1}
    \end{figure*}

\begin{figure*}
   \centering
   \includegraphics[scale=0.124]{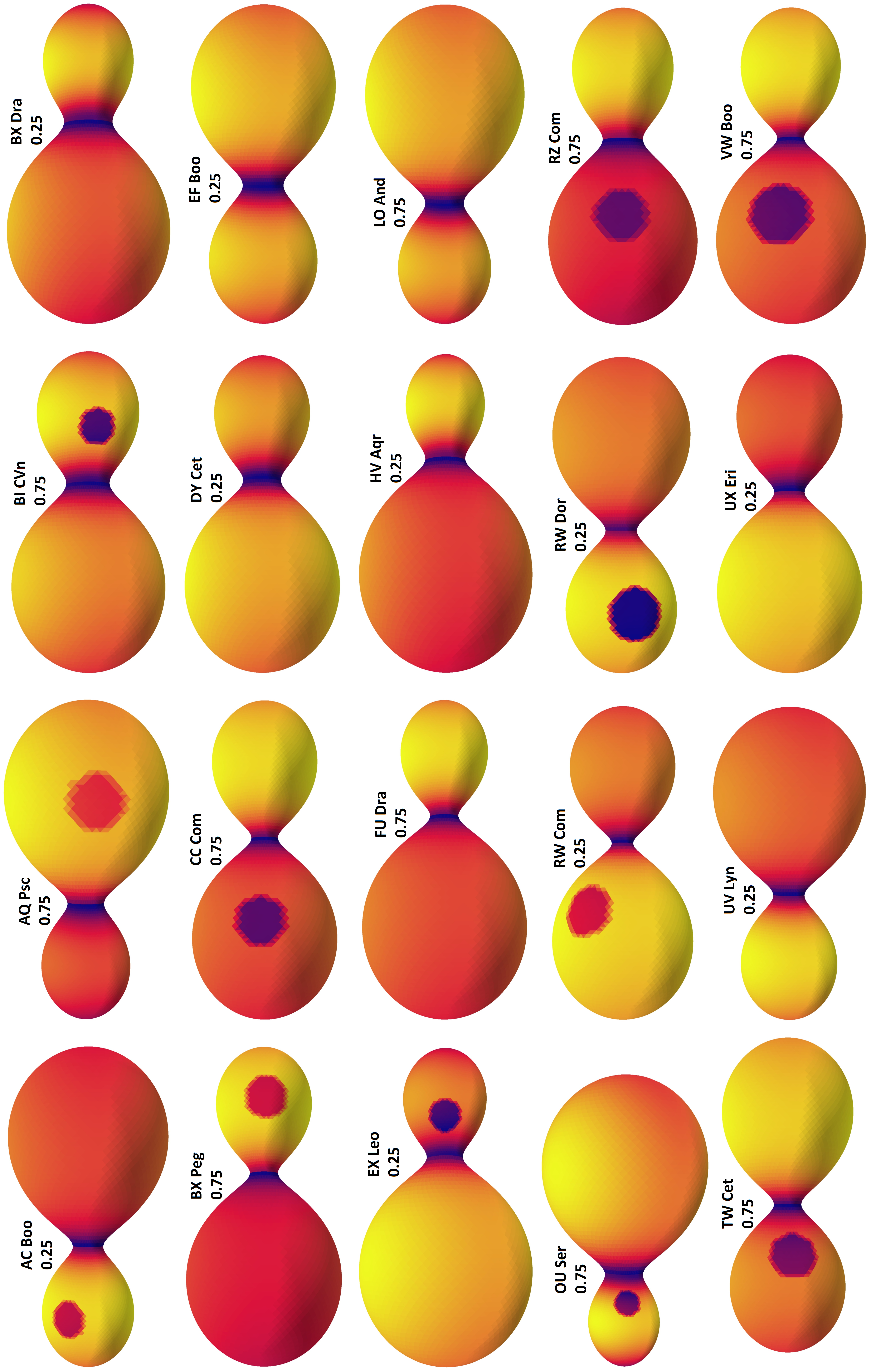}
    \caption{Three-dimensional view of the systems in phases 0.25 or 0.75.}
    \label{Fig2}
    \end{figure*}

The parameters $i$, $q$, $f$, $T_{1,2}$, $\Omega$, $l_1/l_{tot}$, and $r_{(mean)1,2}$ are estimated by modeling the TESS light curves. Table \ref{tab3} presents the light curve analysis results. Figure \ref{Fig1} shows TESS data and synthetic light curves of the binary systems. The geometric structure of the systems in phases 0.25 or 0.75 is shown in Figure \ref{Fig2}. The color in Figure \ref{Fig2} indicates the differences in temperature on the star's surface.

\begin{table*}
\caption{Light curve solutions' results in reference studies of the systems.}
\centering
\begin{center}
\footnotesize
\begin{tabular}{c c c c c c c c}
 \hline
System	& $i^{\circ}$	&	$q=M_2/M_1$	&	$f$	&	$T_1$(K)	&	$T_2$(K)	&	$\Omega_1=\Omega_2$	& Reference\\
\hline
AC Boo	&	86.3(5)	&	3.340(4)	&	0.046	&	6250	&	6241(6)	&	7.034(4)	&	\cite{2010IBVS.5951....1N}[24]	\\
AQ Psc	&	69.06(80)	&	0.266(2)	&	0.35	&	6250(157)	&	6024(150)	&	2.253(12)	&	\cite{2011MNRAS.412.1787D}[30]	\\
BI CVn	&	71.30(10)	&	2.437(4)	&	0.146(11)	&	6125(2)	&	6093	&	5.698(4)	&	\cite{2014NewA...29...57N}[37]	\\
BX Dra	&	80.63(6)	&	0.2884(5)	&	0.515	&	6980	&	6979(2)	&	2.3475(16)	&	\cite{2013PASJ...65....1P}[44]	\\
BX Peg	&	87.2(6)	&	2.66	&	0.171(12)	&	5887(7)	&	5300	&	6.057(8)	&	\cite{2015NewA...41...17L}[49]	\\
CC Com	&	89.8(6)	&	1.90(1)	&	0.17	&	4300	&	4200(60)	&	5.009	&	\cite{2011AN....332..626K}[59]	\\
DY Cet	&	82.48(34)	&	0.356(9)	&	0.24	&	6650(178)	&	6611(176)	&	2.529(5)	&	\cite{2011MNRAS.412.1787D}[30]	\\
EF Boo	&	75.7(2)	&	1.871	&	0.18	&	6450	&	6425(14)	&	4.921(12)	&	\cite{2005AcA....55..123G}[67]	\\
EX Leo	&	60.8(2)	&	0.2	&	0.35	&	6340	&	6110(14)	&	2.186(12)	&	\cite{2010MNRAS.408..464Z}[43]	\\
FU Dra	&	80.4(2)	&	3.989	&	0.190(12)	&	6100	&	5842(6)	&	7.778	&	\cite{2012PASJ...64...48L}[77]	\\
HV Aqr	&	79.186(183)	&	0.145	&	0.569(10)	&	6460	&	6669(7)	&	2.036	&	\cite{2013NewA...21...46L}[83]	\\
LO And	&	80.1(6)	&	0.305(4)	&	0.4	&	6650	&	6690(24)	&	2.401(9)	&	\cite{2015IBVS.6134....1N}[87]	\\
OU Ser	&	50.47(2.24)	&	0.173(17)	&	0.68	&	5940(144)	&	5759(283)	&	2.090(17)	&	\cite{2011MNRAS.412.1787D}[30]	\\
RW Com	&	72.43(29)	&	0.471(6)	&	0.15	&	4830(115)	&	4517(98)	&	5.319(9)	&	\cite{2011MNRAS.412.1787D}[30]	\\
RW Dor	&	77.2(1)	&	1.587	&	0.115(67)	&	5560	&	5287(10)	&	4.64(4)	&	\cite{2019PASJ...71...34S}[105]	\\
RZ Com	&	86.8(6)	&	2.179(9)	&	0.11(1)	&	6276(200)	&	6070	&	5.393(10)	&	\cite{2019IBVS.6266....1N}[111]	\\
TW Cet	&	81.18(10)	&	0.75(3)	&	0.06	&	5865(152)	&	5753(147)	&	3.308(2)	&	\cite{2011MNRAS.412.1787D}[30]	\\
UV Lyn	&	67.6(1)	&	2.685	&	0.18	&	6000	&	5770(5)	&	6.080(1)	&	\cite{2005AcA....55..389Z}[75]	\\
UX Eri	&	75.70(23)	&	0.373(21)	&	0.18	&	6093(153)	&	6006(150)	&	6.065(8)	&	\cite{2011MNRAS.412.1787D}[30]	\\
VW Boo	&	73.81(5)	&	2.336	&	0.108(5)	&	5560	&	5198(3)	&	5.626(3)	&	\cite{2011AJ....141..147L}[130]	\\
\hline
\end{tabular}
\end{center}
\label{tab2}
\end{table*}

\begin{table*}
\caption{The results of the light curve analysis of the systems in this study.}
\centering
\begin{center}
\footnotesize
\begin{tabular}{c c c c c c c c c c c}
 \hline
System	& $i^{\circ}$	&	$q=M_2/M_1$	&	$f$	&	$T_1$(K)	&	$T_2$(K)	&	$\Omega_1=\Omega_2$	&	$l_1/l_{tot}$	&	$r_{(mean)1}$	&	$r_{(mean)2}$	&	Spot\\
\hline
AC Boo	&	$	84.05(50)	$	&	$	3.364(22)	$	&	$	0.199(10)	$	&	$	6378(77)	$	&	$	6091(75)	$	&	6.970(34)	&	0.289(17)	&	0.292(1)	&	0.500(1)	&	1	\\
AQ Psc	&	$	68.60(65)	$	&	$	0.255(27)	$	&	$	0.314(24)	$	&	$	6299(58)	$	&	$	5969(51)	$	&	2.314(60)	&	0.794(9)	&	0.519(2)	&	0.287(2)	&	1	\\
BI CVn	&	$	69.36(36)	$	&	$	2.437(137)	$	&	$	0.356(12)	$	&	$	6186(48)	$	&	$	6010(46)	$	&	5.643(193)	&	0.345(22)	&	0.332(6)	&	0.483(5)	&	1	\\
BX Dra	&	$	79.85(27)	$	&	$	0.259(4)	$	&	$	0.566(20)	$	&	$	6944(46)	$	&	$	7016(44)	$	&	2.281(11)	&	0.744(11)	&	0.531(1)	&	0.304(1)	&	0	\\
BX Peg	&	$	87.30(39)	$	&	$	2.861(20)	$	&	$	0.154(6)	$	&	$	5822(73)	$	&	$	5357(63)	$	&	6.337(30)	&	0.354(21)	&	0.303(1)	&	0.484(1)	&	1	\\
CC Com	&	$	85.83(56)	$	&	$	1.991(46)	$	&	$	0.113(11)	$	&	$	4369(60)	$	&	$	4154(47)	$	&	5.172(71)	&	0.422(7)	&	0.330(2)	&	0.450(2)	&	1	\\
DY Cet	&	$	81.05(65)	$	&	$	0.345(8)	$	&	$	0.287(13)	$	&	$	6666(52)	$	&	$	6569(45)	$	&	2.503(17)	&	0.721(14)	&	0.493(2)	&	0.311(2)	&	0	\\
EF Boo	&	$	74.77(23)	$	&	$	1.882(35)	$	&	$	0.261(9)	$	&	$	6412(48)	$	&	$	6452(48)	$	&	4.929(54)	&	0.363(15)	&	0.347(2)	&	0.456(2)	&	0	\\
EX Leo	&	$	59.22(41)	$	&	$	0.172(26)	$	&	$	0.467(108)	$	&	$	6200(72)	$	&	$	6245(68)	$	&	2.110(71)	&	0.808(35)	&	0.556(11)	&	0.262(12)	&	1	\\
FU Dra	&	$	78.74(29)	$	&	$	3.915(30)	$	&	$	0.164(7)	$	&	$	6090(46)	$	&	$	5848(43)	$	&	7.700(43)	&	0.261(11)	&	0.278(1)	&	0.511(1)	&	0	\\
HV Aqr	&	$	79.35(45)	$	&	$	0.168(2)	$	&	$	0.516(21)	$	&	$	6495(46)	$	&	$	6649(53)	$	&	2.095(6)	&	0.802(10)	&	0.560(1)	&	0.263(1)	&	0	\\
LO And	&	$	79.24(51)	$	&	$	0.304(8)	$	&	$	0.315(13)	$	&	$	6716(53)	$	&	$	6622(62)	$	&	2.415(17)	&	0.741(15)	&	0.505(2)	&	0.301(3)	&	0	\\
OU Ser	&	$	52.82(51)	$	&	$	0.149(37)	$	&	$	0.565(192)	$	&	$	5852(61)	$	&	$	5836(67)	$	&	2.046(105)	&	0.829(46)	&	0.570(18)	&	0.257(21)	&	1	\\
RW Com	&	$	74.14(59)	$	&	$	0.540(80)	$	&	$	0.072(11)	$	&	$	4779(57)	$	&	$	4586(49)	$	&	2.929(153)	&	0.669(51)	&	0.441(14)	&	0.333(13)	&	1	\\
RW Dor	&	$	76.08(30)	$	&	$	1.590(99)	$	&	$	0.144(12)	$	&	$	5506(48)	$	&	$	5318(51)	$	&	4.576(152)	&	0.426(3)	&	0.352(6)	&	0.433(6)	&	1	\\
RZ Com	&	$	85.66(16)	$	&	$	2.348(21)	$	&	$	0.293(8)	$	&	$	6359(60)	$	&	$	6004(54)	$	&	5.559(33)	&	0.373(16)	&	0.330(1)	&	0.476(1)	&	1	\\
TW Cet	&	$	83.62(36)	$	&	$	0.762(56)	$	&	$	0.103(9)	$	&	$	5899(51)	$	&	$	5708(52)	$	&	3.307(97)	&	0.589(3)	&	0.413(6)	&	0.366(6)	&	1	\\
UV Lyn	&	$	65.77(12)	$	&	$	2.696(62)	$	&	$	0.187(7)	$	&	$	5993(34)	$	&	$	5792(38)	$	&	6.096(87)	&	0.323(14)	&	0.310(2)	&	0.481(2)	&	0	\\
UX Eri	&	$	75.62(42)	$	&	$	0.508(50)	$	&	$	0.146(13)	$	&	$	6214(69)	$	&	$	5878(66)	$	&	2.847(97)	&	0.685(35)	&	0.452(9)	&	0.333(9)	&	0	\\
VW Boo	&	$	73.38(13)	$	&	$	2.377(85)	$	&	$	0.127(10)	$	&	$	5504(54)	$	&	$	5267(50)	$	&	5.700(123)	&	0.360(23)	&	0.316(3)	&	0.466(3)	&	1	\\
\hline
\end{tabular}
\end{center}
\label{tab3}
\end{table*}

%%%%%%%%%%%%%%%%%%%%%%%%%%%%%%%%%%%%%%%%%%%%%%%%%%
\vspace{1cm}
\section{Estimation of the Absolute Parameters}
There are various methods to derive absolute parameters, particularly when photometric data is utilized. In some investigations, empirical relationships or the Gaia DR3 parallax method are used to estimate absolute parameters (e.g., \citealt{2015NewA...37...64T}[137], \citealt{2021ApJS..254...10L}[6], \citealt{2023RAA....23i5011P}[138]). Using Gaia DR3 parallax to estimate absolute parameters can have good accuracy, but there are challenges to obtaining proper accuracy. Challenges make this method unsuitable for some binary systems investigations. For example, the parameter that is obtained from the observational data and plays an important role in the calculation process is the maximum apparent magnitude ($V_{max}$). In the first step of the calculation process using Gaia DR3 parallax, $V_{max}$ is used to estimate the absolute magnitude ($M_V$). Therefore, this may not be an appropriate method for calculating the absolute parameters in ground-based observations if the light curve's maxima display more dispersion. On the other hand, the extinction coefficient ($A-V$) should also be reasonable and low value (Table \ref{tab1}). In some cases, the large error for the Gaia DR3 parallax is associated with a large $A_V$ value, although either alone can overshadow the accuracy of the absolute parameter estimates. Large parallax errors usually come in systems with galactic coordinates $b$ between +5 and -5, making it challenging to determine $A_V$ with good accuracy (\citealt{2024PASP..136b4201P}[139]). Also, the initial temperature chosen for the analysis can completely affect the accuracy of the results. The study published by the \cite{2024NewA..11002227P}[140] study discusses the limitations of this method. In this study, due to the lack of a reliable value of $V_max$ for all target systems, we preferred not to use this method.

The mass ratio is one of the most crucial factors in the estimation of the absolute parameters. There are several methods to test or obtain mass ratios from photometric data (\citealt{2021NewA...8601571P}[141], \citealt{2023ApJ...958...84K}[142]). In this study, the initial mass ratio of target systems was obtained using different qualities of spectroscopic data from reference studies. The analysis of light curves showed that the mass ratio did not change significantly compared to the reference studies.

We used the semi-major axis and orbital period ($a-P$) relationship for estimating absolute parameters from the \cite{2024PASP..136b4201P}[139] study (Equation \ref{eq1}). Considering that the uncertainties reported in equation \ref{eq1} have upper and lower limits, we used their average for calculations.

\begin{equation}\label{eq1}
a=(0.372_{\rm-0.114}^{+0.113})+(5.914_{\rm-0.298}^{+0.272})\times P
\end{equation}

Then, using the well-known equation of Kepler's third law (Equation \ref{eq2}), we obtained the total mass ($M_1+M_2$) of the components. Since $q=M_2/M_1$, the values of each star's mass were estimated.

\begin{equation}\label{eq2}
\frac{a^3}{G(M_1+M_2)}=\frac{P^2}{4\pi^2}
\end{equation}

The values of $r_{mean}$ obtained from light curve solutions (Table \ref{tab3}) and the radius ($R$) of each star were calculated (Equation \ref{eq3}).

\begin{equation}\label{eq3}
a=\frac{R}{r}
\end{equation}

According to the effective temperature obtained from the light curve analysis and the calculated radius, the luminosity ($L$) was calculated (Equation \ref{eq4}).

\begin{equation}\label{eq4}
L=4\pi R^2 \sigma T^{4}
\end{equation}

The absolute bolometric magnitude of stars was calculated and $M_{bol\odot}$ considered 4.73 from the \cite{2010AJ....140.1158T}[143] throughout this estimating process (Equation \ref{eq5}).

\begin{equation}\label{eq5}
M_{bol}-M_{bol_{\odot}}=-2.5log(\frac{L}{L_{\odot}})
\end{equation}

The surface gravity of each star was also calculated based on its mass and radius (Equation \ref{eq6}).

\begin{equation}\label{eq6}
g=G_{\odot}(\frac{M}{R^2})
\end{equation}

The uncertainties of the absolute parameters were calculated using the errors determined by the PHOEBE code for the light curve elements used in the process, such as $T_{1,2}$, $r_{mean1,2}$, and $q$. Table \ref{tab4} contains the results of the estimated absolute parameters.

\begin{table*}
\centering
\caption{Estimated absolute parameters of 20 contact binary systems using Gaia DR3 parallax.}
\begin{center}
\footnotesize
\begin{tabular}{c c c c c c c c c c c c}
 \hline
System	&	$M_1(M_\odot)$	&	$M_2(M_\odot)$	&	$R_1(R_\odot)$	&	$R_2(R_\odot)$	&	$L_1(L_\odot)$	&	$L_2(L_\odot)$	&	$M_{bol1}$	&	$M_{bol2}$	&	$log(g)_1$	&	$log(g)_2$	&	$a(R_\odot)$\\
\hline
AC Boo	&	0.37(10)	&	1.23(35)	&	0.72(7)	&	1.23(11)	&	0.77(19)	&	1.87(46)	&	5.02(24)	&	4.05(24)	&	4.29(3)	&	4.35(4)	&	2.46(21)	\\
AQ Psc	&	1.53(35)	&	0.39(14)	&	1.65(14)	&	0.91(8)	&	3.88(84)	&	0.96(21)	&	3.26(21)	&	4.78(22)	&	4.19(2)	&	4.11(6)	&	3.18(25)	\\
BI CVn	&	0.49(11)	&	1.19(35)	&	0.88(9)	&	1.28(12)	&	1.02(26)	&	1.92(46)	&	4.71(25)	&	4.02(23)	&	4.24(1)	&	4.30(3)	&	2.64(22)	\\
BX Dra	&	1.74(41)	&	0.45(11)	&	2.02(15)	&	1.15(9)	&	8.52(60)	&	2.91(55)	&	2.40(19)	&	3.57(19)	&	4.07(3)	&	3.97(3)	&	3.80(28)	\\
BX Peg	&	0.37(11)	&	1.06(34)	&	0.62(6)	&	0.98(10)	&	0.39(11)	&	0.72(19)	&	5.75(26)	&	5.09(25)	&	4.43(3)	&	4.48(4)	&	2.03(19)	\\
CC Com	&	0.43(14)	&	0.87(31)	&	0.55(6)	&	0.75(8)	&	0.10(3)	&	0.15(4)	&	7.22(29)	&	6.77(28)	&	4.59(3)	&	4.62(4)	&	1.68(18)	\\
DY Cet	&	1.36(34)	&	0.47(13)	&	1.47(12)	&	0.93(8)	&	3.84(82)	&	1.44(31)	&	3.27(21)	&	4.33(21)	&	4.24(3)	&	4.17(4)	&	2.98(24)	\\
EF Boo	&	0.62(15)	&	1.16(32)	&	0.99(9)	&	1.30(11)	&	1.50(33)	&	2.66(57)	&	4.29(22)	&	3.67(21)	&	4.23(2)	&	4.27(3)	&	2.86(23)	\\
EX Leo	&	1.49(36)	&	0.26(11)	&	1.55(16)	&	0.73(10)	&	3.20(89)	&	0.73(25)	&	3.47(27)	&	5.07(32)	&	4.23(1)	&	4.12(5)	&	2.79(23)	\\
FU Dra	&	0.30(9)	&	1.19(36)	&	0.61(6)	&	1.12(11)	&	0.46(11)	&	1.32(31)	&	5.58(23)	&	4.43(23)	&	4.35(3)	&	4.42(4)	&	2.19(20)	\\
HV Aqr	&	1.42(39)	&	0.24(7)	&	1.45(13)	&	0.68(6)	&	3.37(73)	&	0.82(18)	&	3.41(21)	&	4.95(22)	&	4.27(3)	&	4.15(4)	&	2.59(22)	\\
LO And	&	1.28(34)	&	0.39(12)	&	1.32(12)	&	0.79(8)	&	3.22(72)	&	1.08(27)	&	3.46(22)	&	4.65(24)	&	4.30(3)	&	4.23(4)	&	2.62(22)	\\
OU Ser	&	1.28(34)	&	0.19(11)	&	1.21(16)	&	0.55(10)	&	1.55(51)	&	0.31(15)	&	4.25(31)	&	5.99(41)	&	4.38(1)	&	4.24(5)	&	2.13(20)	\\
RW Com	&	0.87(24)	&	0.47(22)	&	0.78(11)	&	0.59(9)	&	0.29(10)	&	0.14(5)	&	6.08(33)	&	6.87(34)	&	4.59(1)	&	4.56(5)	&	1.78(18)	\\
RW Dor	&	0.56(15)	&	0.88(30)	&	0.73(8)	&	0.89(10)	&	0.44(12)	&	0.57(16)	&	5.63(27)	&	5.33(27)	&	4.46(1)	&	4.48(4)	&	2.06(20)	\\
RZ Com	&	0.47(13)	&	1.10(32)	&	0.78(7)	&	1.13(10)	&	0.90(22)	&	1.50(35)	&	4.84(23)	&	4.29(23)	&	4.32(3)	&	4.37(4)	&	2.37(21)	\\
TW Cet	&	0.86(22)	&	0.66(23)	&	0.93(10)	&	0.82(9)	&	0.94(25)	&	0.65(18)	&	4.80(26)	&	5.20(26)	&	4.44(1)	&	4.42(4)	&	2.25(20)	\\
UV Lyn	&	0.48(12)	&	1.28(35)	&	0.88(8)	&	1.36(12)	&	0.89(19)	&	1.87(40)	&	4.85(21)	&	4.05(21)	&	4.23(2)	&	4.28(3)	&	2.83(23)	\\
UX Eri	&	1.22(27)	&	0.62(21)	&	1.36(14)	&	1.00(11)	&	2.48(67)	&	1.08(31)	&	3.74(26)	&	4.65(27)	&	4.26(1)	&	4.23(4)	&	3.01(24)	\\
VW Boo	&	0.47(12)	&	1.11(34)	&	0.76(7)	&	1.12(11)	&	0.47(12)	&	0.87(21)	&	5.54(25)	&	4.89(24)	&	4.35(2)	&	4.39(4)	&	2.40(21)	\\
\hline
\end{tabular}
\end{center}
\label{tab4}
\end{table*}

%%%%%%%%%%%%%%%%%%%%%%%%%%%%%%%%%%%%%%%%%%%%%%%%%%
\vspace{1cm}
\section{Discussion and Conclusion}
We selected 20 contact binary systems and one of the latest published studies for each of them. The target systems have or used spectroscopic results. Except for DY Cet, none of them have used TESS data for analysis. Quality photometric data, like space-based data, is important for obtaining accurate light curve parameters and then estimating absolute parameters.

We conducted the light curve analysis using the PHOEBE Python code and TESS observations. We considered the effective temperature and mass ratio reported in the studies as input values.
The results of the light curve analysis showed that the mass ratio of the systems has changed slightly. The minimum difference between the mass ratio of our results and reference studies is related to the BI CVn system with 0\%, and the maximum difference is for the BX Peg system with 7\%. The elapsed times from the reference studies and precise TESS data rather than ground-based observations can account for some of the discrepancies in the light curve analysis results.

The stars in contact binary systems have a small temperature difference due to mass and energy transfer (\citealt{2021ApJS..254...10L}[6]). The difference in effective temperature between the two stars in the BX Peg system had the maximum value at 465 K among the target systems, while the OU Ser system with 16 K had the lowest. Table \ref{tab5} lists the temperature difference between companions. Based on the stars' effective temperatures, we estimated the spectral type of the stars using \cite{2000asqu.book.....C}[144] study (Table \ref{tab5}). We also checked the results of the effective temperature in this study with the reports of Gaia (\citealt{2023AA...674A..33G}[145]) and TESS, so that the difference is less than 5\%.

Additionally, EX Leo, HV Aqr, and OU Ser are low mass ratio contact systems with values of 0.172(28), 0.168(23), and 0.149(24), respectively. Although these three systems have a low mass ratio, they are not close to the extremely low mass ratio cutoff, which is less than 0.1 (\citealt{2021ApJ...922..122L}[146]).
On the other hand, we have three systems (BX Dra, HV Aqr, OU Ser) with fillout factors greater than 50\% and mass ratios less than 0.25. These specifications relate to deep overcontact binaries suggested by the \cite{2005AJ....130..224Q}[147] study, which found that this type of star is likely to be the progenitor of a blue straggler/FK Com-type star (\citealt{2009AJ....138..540Y}[148], \citealt{2011AJ....141..151Q}[149], \citealt{liao2017}[150], \citealt{2021ApJ...922..122L}[146]).

We estimated the absolute parameters using the semi-major axis and orbital period relationship. So, some values from the light curve solutions ($T_{1,2}$, $q$, $r_{mean1,2}$), and the orbital period used in the calculation process.

\begin{figure*}
   \centering
   \includegraphics[width=\textwidth]{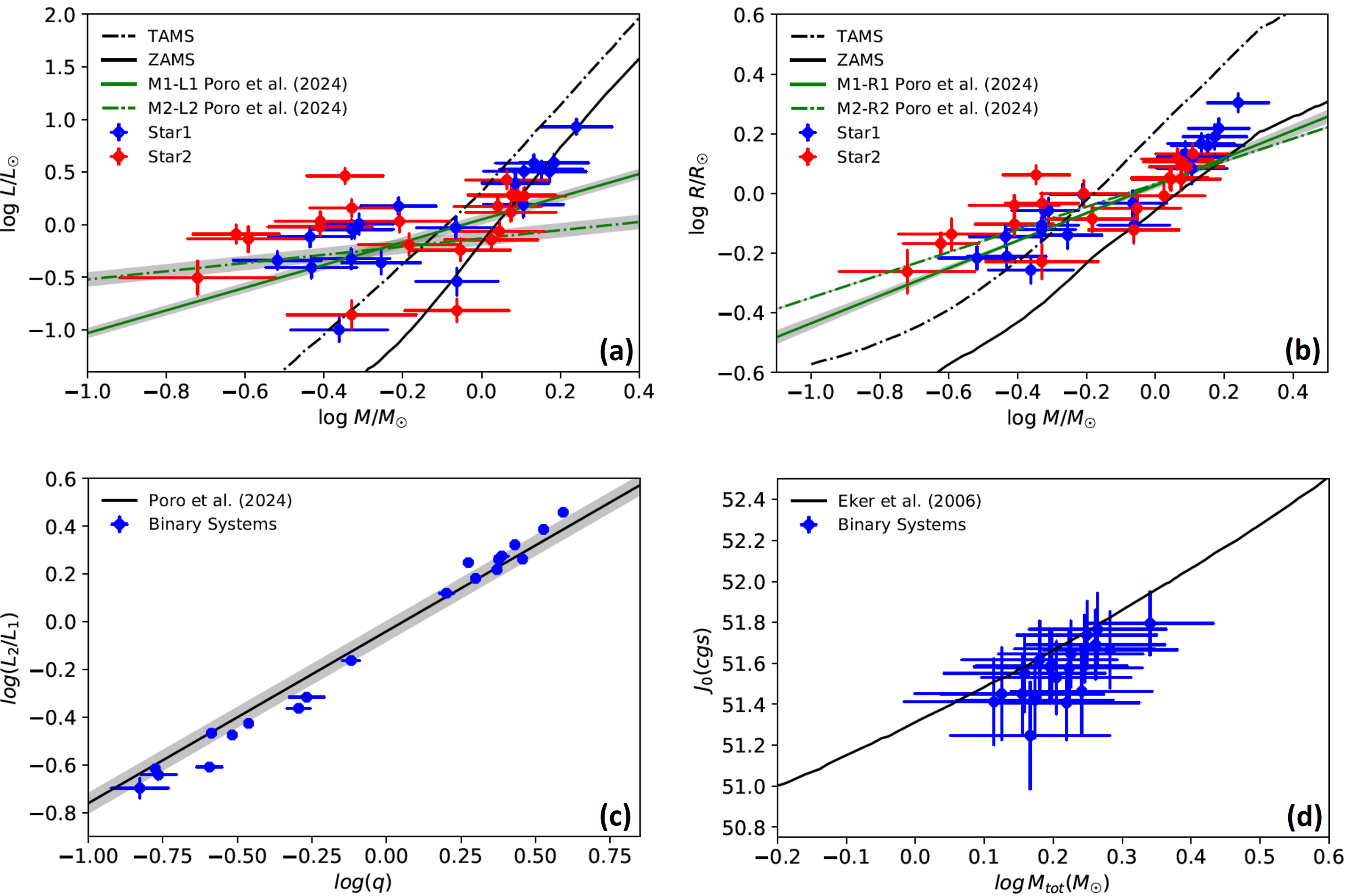}
      \caption{(a) $M_{1,2}-L_{1,2}$, (b) $M_{1,2}-R_{1,2}$, (c) $q-L_{ratio}$, and (d) $M_{tot}-J_0$ diagrams.}
    \label{Fig3}
    \end{figure*}

The positions of the stars were displayed using the Zero-Age Main Sequence (ZAMS) and the Terminal-Age Main Sequence (TAMS) on the Mass-Luminosity ($M-L$) and Mass-Radius ($M-R$) diagrams, based on the absolute parameters (Figure \ref{Fig3}a,b). Additionally, the outcomes have been compared with the theoretical fits from the \cite{2024RAA....24a5002P}[10] study, and as expected the $M-L$ and $M-R$ connections are weak.
Figure \ref{Fig3}c shows the position of the systems compared to the $q-L_{ratio}$ theoretical fit obtained from the \cite{2024RAA....24a5002P}[10] study, with which they are in good agreement. We utilized the following equation from the \cite{2006MNRAS.373.1483E}[151] study to calculate the orbital angular momentum ($J_0$) of the systems:

\begin{equation}\label{eq7}
J_0=\frac{q}{(1+q)^2} \sqrt[3] {\frac{G^2}{2\pi}M^5P}
\end{equation}

\noindent where $q$ is the mass ratio, $M$ is the total mass of the system, $P$ is the orbital period, and $G$ is the gravitational constant. The results of estimating the $J_0$ are listed in Table \ref{tab5} along with the total mass of the systems. We also determined the subtype of each system (Table \ref{tab5}). Therefore, we considered the systems where the more massive star has a hotter effective temperature than the companion as A-type and otherwise as W-type. Additionally, the target systems' position is depicted in the $logM_{tot}-logJ_0$ diagram (Figure \ref{Fig3}d), which indicates that they are in a contact binary systems region.

\begin{figure}
   \centering
   \includegraphics[scale=0.5]{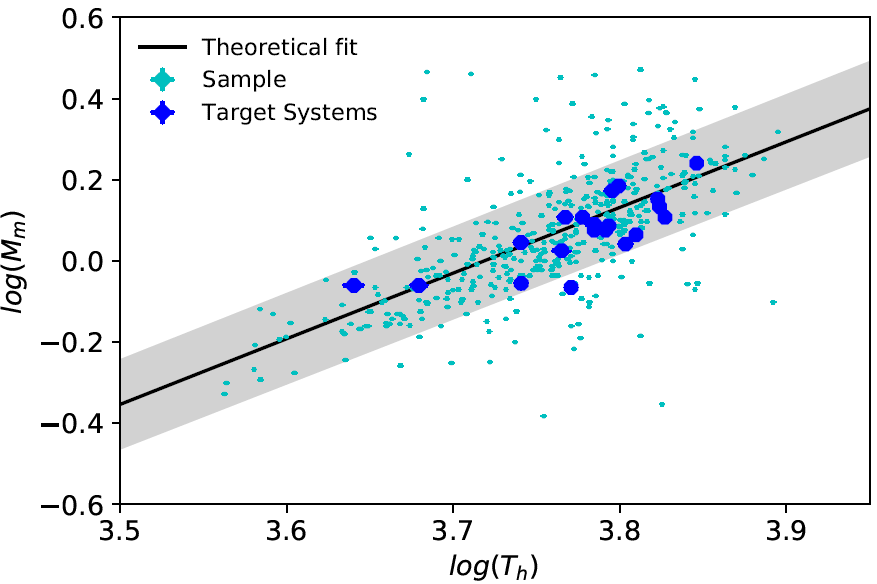}
      \caption{The diagram of the relationship between the effective temperature ($T_h$) and the mass ($M_m$) of the primary component in contact binary stars in which the studied systems are displayed.}
    \label{Fig4}
    \end{figure}

The temperature-mass relationship for contact binary systems was presented by \cite{2024RAA....24e5001P}[152] with a linear fit. They made use of 428 contact systems from the \cite{2021ApJS..254...10L}[6] study's sample. The hotter component ($T_h$) and the mass of the more massive star ($M_m$) were considered for this relation by \cite{2024RAA....24e5001P}[152]. Our target systems are positioned on the $T_h-M_m$ diagram (Figure \ref{Fig4}), which indicates good agreement with the theoretical fit and uncertainty.

\begin{table*}
\centering
\caption{The spectral type (Sp.) of stars, temperature difference of companions, total mass, orbital angular momentum, and subtype of the target binary systems.}
\begin{center}
\footnotesize
\begin{tabular}{c c c c c c c}
 \hline
System & Sp. Star1 & Sp. Star2 & $\Delta T(K)$ & $M_{tot}(M_{\odot})$ & $J_0$ & Subtype\\
\hline
AC Boo & F5	&	F8	&	287	&	1.60(46)	&	51.53(18) 	&	W	\\
AQ Psc & F6	&	G1	&	330	&	1.92(49)	&	51.67(19) 	&	A	\\
BI CVn & F8	&	G0	&	176	&	1.68(46)	&	51.65(17) 	&	W	\\
BX Dra & F1	&	F1	&	72	&	2.19(52)	&	51.80(16) 	&	W	\\
BX Peg & G3	&	K0	&	465	&	1.43(45)	&	51.45(20) 	&	W	\\
CC Com & K5	&	K5	&	215	&	1.30(46)	&	51.41(21) 	&	W	\\
DY Cet & F3	&	F3	&	97	&	1.83(48)	&	51.69(17) 	&	A	\\
EF Boo & F5	&	F5	&	40	&	1.77(47)	&	51.74(17) 	&	A	\\
EX Leo & F7	&	F7	&	45	&	1.74(47)	&	51.46(21) 	&	W	\\
FU Dra & G0	&	G3	&	242	&	1.49(45)	&	51.42(19) 	&	W	\\
HV Aqr & F5	&	F3	&	154	&	1.66(46)	&	51.41(18) 	&	W	\\
LO And & F2	&	F3	&	94	&	1.67(46)	&	51.58(18) 	&	A	\\
OU Ser & G3	&	G3	&	16	&	1.47(45)	&	51.25(26) 	&	A	\\
RW Com & K3	&	K3	&	193	&	1.33(45)	&	51.45(23) 	&	A	\\
RW Dor & G8	&	K0	&	188	&	1.44(45)	&	51.55(19) 	&	W	\\
RZ Com & F5	&	G0	&	355	&	1.57(45)	&	51.58(18) 	&	W	\\
TW Cet & G2	&	G6	&	191	&	1.51(45)	&	51.62(19) 	&	A	\\
UV Lyn & G0	&	G3	&	201	&	1.76(47)	&	51.67(17) 	&	W	\\
UX Eri & F7	&	G2	&	336	&	1.84(48)	&	51.77(18) 	&	A	\\
VW Boo & G8	&	K0	&	237	&	1.58(46)	&	51.59(18) 	&	W	\\
\hline
\end{tabular}
\end{center}
\label{tab5}
\end{table*}

%%%%%%%%%%%%%%%%%%%%%%%%%%%%%%%%%%%%%%%%%%%%%%%%%%
\vspace{1cm}
\section*{Data availability}
We used TESS photometric data in this study, and they are available at the MAST portal.

%%%%%%%%%%%%%%%%%%%%%%%%%%%%%%%%%%%%%%%%%%%%%%%%%%
\vspace{1cm}
\section*{Acknowledgements}
This manuscript was prepared by the BSN project (\url{https://bsnp.info/}). We used data from the European Space Agency's mission Gaia (\url{http://www.cosmos.esa.int/gaia}). We made use of the SIMBAD database, which is operated by CDS in Strasbourg, France (\url{http://simbad.u-strasbg.fr/simbad/}). This work includes data from the TESS mission observations. Funding for the TESS mission is provided by the NASA Explorer Program. We acknowledge the TESS team for its support of this study.

%%%%%%%%%%%%%%%%%%%%%%%%%%%%%%%%%%%%%%%%%%%%%%%%%%
\vspace{1cm}
\section*{ORCID iDs}
\noindent Ehsan Paki: 0000-0001-9746-2284\\
Atila Poro: 0000-0002-0196-9732\\

%%%%%%%%%%%%%%%%%%%%%%%%%%%%%%%%%%%%%%%%%%%%%%%%%%


\vspace{1cm}
\begin{thebibliography}{}
\bibitem[\protect\citeauthoryear{Kopal}{1959}]{1959cbs..book.....K} [1] Kopal Z., 1959, cbs..book
\bibitem[\protect\citeauthoryear{Kuiper}{1941}]{kuiper1941} [2] Kuiper G.P., 1941, ApJ, 93, 133
\bibitem[\protect\citeauthoryear{Yakut \& Eggleton}{2005}]{2005ApJ...629.1055Y} [3] Yakut K., Eggleton P.~P., 2005, ApJ, 629, 1055. doi:10.1086/431300
\bibitem[\protect\citeauthoryear{Binnendijk}{1970}]{binnendijk1970} [4] Binnendijk L., 1970, vistas in Astronomy, 12, 217
\bibitem[\protect\citeauthoryear{Lucy \& Wilson}{1979}]{1979ApJ...231..502L} [5] Lucy L.~B., Wilson R.~E., 1979, ApJ, 231, 502. doi:10.1086/157212
\bibitem[\protect\citeauthoryear{Latkovi{\'c}, {\v{C}}eki, \& Lazarevi{\'c}}{2021}]{2021ApJS..254...10L} [6] Latkovi{\'c} O., {\v{C}}eki A., Lazarevi{\'c} S., 2021, ApJS, 254, 10. doi:10.3847/1538-4365/abeb23
\bibitem[\protect\citeauthoryear{Loukaidou et al.}{2022}]{2022MNRAS.514.5528L} [7] Loukaidou G.~A., Gazeas K.~D., Palafouta S., Athanasopoulos D., Zola S., Liakos A., Niarchos P.~G., et al., 2022, MNRAS, 514, 5528. doi:10.1093/mnras/stab3424
\bibitem[\protect\citeauthoryear{Zhang \& Qian}{2020}]{2020MNRAS.497.3493Z} [8] Zhang X.-D., Qian S.-B., 2020, MNRAS, 497, 3493. doi:10.1093/mnras/staa2166
\bibitem[\protect\citeauthoryear{O'Connell}{1951}]{1951PRCO....2...85O} [9] O'Connell D.~J.~K., 1951, PRCO, 2, 85
\bibitem[\protect\citeauthoryear{Poro et al.}{2024}]{2024RAA....24a5002P} [10] Poro A., Paki E., Alizadehsabegh A., Khodadadilori M., Salehian S.~R., Hedayatjoo M., Hashemi F., et al., 2024, RAA, 24, 015002. doi:10.1088/1674-4527/ad0866
\bibitem[\protect\citeauthoryear{Li et al.}{2021}]{2021AJ....162...13L} [11] Li K., Xia Q.-Q., Kim C.-H., Gao X., Hu S.-M., Guo D.-F., Gao D.-Y., et al., 2021, AJ, 162, 13. doi:10.3847/1538-3881/abfc53
\bibitem[\protect\citeauthoryear{Y{\i}ld{\i}r{\i}m}{2022}]{2022RAA....22e5013Y} [12] Y{\i}ld{\i}r{\i}m M.~F., 2022, RAA, 22, 055013. doi:10.1088/1674-4527/ac5ee8
\bibitem[\protect\citeauthoryear{Ricker et al.}{2010}]{2010AAS...21545006R} [13] Ricker G.~R., Latham D.~W., Vanderspek R.~K., Ennico K.~A., Bakos G., Brown T.~M., Burgasser A.~J., et al., 2010, AAS
\bibitem[\protect\citeauthoryear{Stassun et al.}{2018}]{2018AJ....156..102S} [14] Stassun K.~G., Oelkers R.~J., Pepper J., Paegert M., De Lee N., Torres G., Latham D.~W., et al., 2018, AJ, 156, 102. doi:10.3847/1538-3881/aad050
\bibitem[\protect\citeauthoryear{Schlafly \& Finkbeiner}{2011}]{2011ApJ...737..103S} [15] Schlafly E.~F., Finkbeiner D.~P., 2011, ApJ, 737, 103. doi:10.1088/0004-637X/737/2/103
\bibitem[\protect\citeauthoryear{Green et al.}{2019}]{2019ApJ...887...93G} [16] Green G.~M., Schlafly E., Zucker C., Speagle J.~S., Finkbeiner D., 2019, ApJ, 887, 93. doi:10.3847/1538-4357/ab5362
\bibitem[\protect\citeauthoryear{Herz}{1995}]{herz1995marcus} [17] Herz, I Marcus, 1995, Identity Or History?: Marcus Herz and the End of the Enlightenment, Wayne State University Press, p.300
\bibitem[\protect\citeauthoryear{Zessewitsch}{1956}]{zessewitsch1956} [18] Zessewitsch W.P., 1956, AZ Kasan, v.171
\bibitem[\protect\citeauthoryear{Mauder}{1964}]{1964ZA.....60..222M} [19] Mauder H., 1964, ZA, 60, 222
\bibitem[\protect\citeauthoryear{Mancuso \& Milano}{1974}]{1974IBVS..904....1M} [20] Mancuso S., Milano L., 1974, IBVS, 904, 1
\bibitem[\protect\citeauthoryear{Mancuso, Milano, \& Russo}{1977}]{1977AAS...29...57M} [21] Mancuso S., Milano L., Russo G., 1977, A\&AS, 29, 57
\bibitem[\protect\citeauthoryear{Mancuso, Milano, \& Russo}{1978}]{1978AA....63..193M} [22] Mancuso S., Milano L., Russo G., 1978, A\&A, 63, 193
\bibitem[\protect\citeauthoryear{Schieven et al.}{1983}]{1983AAS...52..463S} [23] Schieven G., Morton J.~C., McLean B.~J., Hughes V.~A., 1983, A\&AS, 52, 463
\bibitem[\protect\citeauthoryear{Nelson}{2010}]{2010IBVS.5951....1N} [24] Nelson R.~H., 2010, IBVS, 5951, 1
\bibitem[\protect\citeauthoryear{Sarma \& Radharkrishnan}{1982}]{1982IBVS.2073....1S} [25] Sarma M.~B.~K., Radharkrishnan K.~R., 1982, IBVS, 2073, 1
\bibitem[\protect\citeauthoryear{Lu \& Rucinski}{1999}]{1999AJ....118..515L} [26] Lu W., Rucinski S.~M., 1999, AJ, 118, 515. doi:10.1086/300933
\bibitem[\protect\citeauthoryear{Yamasaki}{2005}]{2005ApSS.296..277Y} [27] Yamasaki A., 2005, Ap\&SS, 296, 277. doi:10.1007/s10509-005-4830-3
\bibitem[\protect\citeauthoryear{Pribulla \& Rucinski}{2006}]{2006AJ....131.2986P} [28] Pribulla T., Rucinski S.~M., 2006, AJ, 131, 2986. doi:10.1086/503871
\bibitem[\protect\citeauthoryear{Djura{\v{s}}evi{\'c} et al.}{2006}]{2006PASA...23..154D} [29] Djura{\v{s}}evi{\'c} G., Dimitrov D., Arbutina B., Albayrak B., Selam S.~O., Atanackovi{\'c}-Vukmanovi{\'c} O., 2006, PASA, 23, 154. doi:10.1071/AS06016
\bibitem[\protect\citeauthoryear{Deb \& Singh}{2011}]{2011MNRAS.412.1787D} [30] Deb S., Singh H.~P., 2011, MNRAS, 412, 1787. doi:10.1111/j.1365-2966.2010.18016.x
\bibitem[\protect\citeauthoryear{Kippenhahn}{1955}]{kippenhahn1955} [31] Kippenhahn R., 1955, VeBab, v.11
\bibitem[\protect\citeauthoryear{Filatov}{1960}]{1960ATsir.215...20F} [32] Filatov G.~S., 1960, ATsir, 215
\bibitem[\protect\citeauthoryear{Demircan}{1987}]{1987ApSS.135..169D} [33] Demircan O., 1987, Ap\&SS, 135, 169. doi:10.1007/BF00644471
\bibitem[\protect\citeauthoryear{Lu}{1988}]{1988AcApS...8..259L} [34] Lu W.-X., 1988, AcApS, 8, 259
\bibitem[\protect\citeauthoryear{Maceroni \& van't Veer}{1996}]{1996AA...311..523M} [35] Maceroni C., van't Veer F., 1996, A\&A, 311, 523
\bibitem[\protect\citeauthoryear{Qian et al.}{2008}]{2008AJ....136.2493Q} [36] Qian S.-B., He J.-J., Liu L., Zhu L.-Y., Liao W.~P., 2008, AJ, 136, 2493. doi:10.1088/0004-6256/136/6/2493
\bibitem[\protect\citeauthoryear{Nelson et al.}{2014}]{2014NewA...29...57N} [37] Nelson R.~H., {\c{S}}enavci H.~V., Ba{\c{s}}t{\"u}rk {\"O}., Bahar E., 2014, NewA, 29, 57. doi:10.1016/j.newast.2013.11.006
\bibitem[\protect\citeauthoryear{Strohmeier}{1958}]{strohmeier1958} [38] Strohmeier W., 1958, Kl. Veroffent. Sternwarte Bamberg, v.24
\bibitem[\protect\citeauthoryear{Smith}{1990}]{1990PASP..102..124S} [39] Smith H.~A., 1990, PASP, 102, 124. doi:10.1086/132617
\bibitem[\protect\citeauthoryear{Pych et al.}{2004}]{2004AJ....127.1712P} [40] Pych W., Rucinski S.~M., DeBond H., Thomson J.~R., Capobianco C.~C., Blake R.~M., Og{\l}oza W., et al., 2004, AJ, 127, 1712. doi:10.1086/382105
\bibitem[\protect\citeauthoryear{S{\'a}nchez-Bajo et al.}{2007}]{2007ApSS.312..151S} [41] S{\'a}nchez-Bajo F., Garc{\'\i}a-Melendo E., G{\'o}mez-Forrellad J.~M., 2007, Ap\&SS, 312, 151. doi:10.1007/s10509-007-9668-4
\bibitem[\protect\citeauthoryear{Kim et al.}{2009}]{2009IBVS.5872....1K} [42] Kim D., Jeon Y.-B., Lee U., Jang H.-E., Cho S., Park Y.-H., 2009, IBVS, 5872, 1
\bibitem[\protect\citeauthoryear{Zola et al.}{2010}]{2010MNRAS.408..464Z} [43] Zola S., Gazeas K., Kreiner J.~M., Ogloza W., Siwak M., Koziel-Wierzbowska D., Winiarski M., 2010, MNRAS, 408, 464. doi:10.1111/j.1365-2966.2010.17129.x
\bibitem[\protect\citeauthoryear{Park et al.}{2013}]{2013PASJ...65....1P} [44] Park J.-H., Lee J.~W., Kim S.-L., Lee C.-U., Jeon Y.-B., 2013, PASJ, 65, 1. doi:10.1093/pasj/65.1.1
\bibitem[\protect\citeauthoryear{Shapley \& Hughes}{1934}]{shapley1934variable} [45] Shapley, H. and Hughes, E.M., 1934, Harvard Ann., v.90
\bibitem[\protect\citeauthoryear{Samec \& Hube}{1991}]{1991AJ....102.1171S} [46] Samec R.~G., Hube D.~P., 1991, AJ, 102, 1171. doi:10.1086/115943
\bibitem[\protect\citeauthoryear{Lee et al.}{2009}]{2009PASP..121.1366L} [47] Lee J.~W., Kim S.-L., Lee C.-U., Youn J.-H., 2009, PASP, 121, 1366. doi:10.1086/649230
\bibitem[\protect\citeauthoryear{Alton}{2013}]{2013JAVSO..41..227A} [48] Alton K.~B., 2013, JAVSO, 41, 227
\bibitem[\protect\citeauthoryear{Li et al.}{2015}]{2015NewA...41...17L} [49] Li K., Hu S., Guo D., Jiang Y., Gao D., Chen X., 2015, NewA, 41, 17. doi:10.1016/j.newast.2015.04.010
\bibitem[\protect\citeauthoryear{Hoffmeister}{1964}]{1964AN....288...49H} [50] Hoffmeister C., 1964, AN, 288, 49. doi:10.1002/asna.19652880402
\bibitem[\protect\citeauthoryear{Rucinski}{1976}]{1976PASP...88..777R} [51] Rucinski S.~M., 1976, PASP, 88, 777. doi:10.1086/130024
\bibitem[\protect\citeauthoryear{Breinhorst \& Hoffmann}{1982}]{1982ApSS..86..107B} [52] Breinhorst R.~A., Hoffmann M., 1982, Ap\&SS, 86, 107. doi:10.1007/BF00651833
\bibitem[\protect\citeauthoryear{Bradstreet}{1985}]{1985ApJS...58..413B} [53] Bradstreet D.~H., 1985, ApJS, 58, 413. doi:10.1086/191047
\bibitem[\protect\citeauthoryear{Zhou}{1988}]{1988ApSS.141..199Z} [54] Zhou H.-N., 1988, Ap\&SS, 141, 199. doi:10.1007/BF00639488
\bibitem[\protect\citeauthoryear{Linnell \& Olson}{1989}]{1989ApJ...343..909L} [55] Linnell A.~P., Olson E.~C., 1989, ApJ, 343, 909. doi:10.1086/167760
\bibitem[\protect\citeauthoryear{Rucinski, Whelan, \& Worden}{1977}]{1977PASP...89..684R} [56] Rucinski S.~M., Whelan J.~A.~J., Worden S.~P., 1977, PASP, 89, 684. doi:10.1086/130208
\bibitem[\protect\citeauthoryear{McLean \& Hilditch}{1983}]{1983MNRAS.203....1M} [57] McLean B.~J., Hilditch R.~W., 1983, MNRAS, 203, 1. doi:10.1093/mnras/203.1.1
\bibitem[\protect\citeauthoryear{Pribulla et al.}{2007}]{2007AJ....133.1977P} [58] Pribulla T., Rucinski S.~M., Conidis G., DeBond H., Thomson J.~R., Gazeas K., Og{\l}oza W., 2007, AJ, 133, 1977. doi:10.1086/512772
\bibitem[\protect\citeauthoryear{K{\"o}se et al.}{2011}]{2011AN....332..626K} [59] K{\"o}se O., Kalomeni B., Keskin V., Ula{\c{s}} B., Yakut K., 2011, AN, 332, 626. doi:10.1002/asna.201011566
\bibitem[\protect\citeauthoryear{Perryman et al.}{1997}]{1997AA...323L..49P} [60] Perryman M.~A.~C., Lindegren L., Kovalevsky J., Hoeg E., Bastian U., Bernacca P.~L., Cr{\'e}z{\'e} M., et al., 1997, A\&A, 323, L49
\bibitem[\protect\citeauthoryear{Selam}{2004}]{2004AA...416.1097S} [61] Selam S.~O., 2004, A\&A, 416, 1097. doi:10.1051/0004-6361:20034578
\bibitem[\protect\citeauthoryear{Pribulla et al.}{2009}]{2009AJ....137.3655P} [62] Pribulla T., Rucinski S.~M., Blake R.~M., Lu W., Thomson J.~R., DeBond H., Karmo T., et al., 2009, AJ, 137, 3655. doi:10.1088/0004-6256/137/3/3655
\bibitem[\protect\citeauthoryear{Samec et al.}{1999}]{1999IBVS.4811....1S} [63] Samec R.~G., Tuttle J.~P., Brougher J.~A., Moore J.~E., Faulkner D.~R., 1999, IBVS, 4811, 1
\bibitem[\protect\citeauthoryear{Rucinski et al.}{2001}]{2001AJ....122.1974R} [64] Rucinski S.~M., Lu W., Mochnacki S.~W., Og{\l}oza W., Stachowski G., 2001, AJ, 122, 1974. doi:10.1086/323106
\bibitem[\protect\citeauthoryear{Gothard, Van Hamme, \& Samec}{2000}]{2000AAS...197.4803G} [65] Gothard N.~W., Van Hamme W., Samec R.~G., 2000, AAS
\bibitem[\protect\citeauthoryear{Ozdemir et al.}{2001}]{2001IBVS.5033....1O} [66] Ozdemir S., Demircan O., Erdem A., Cicek C., Bulut I., Soydugan F., Soydugan E., 2001, IBVS, 5033, 1
\bibitem[\protect\citeauthoryear{Gazeas et al.}{2005}]{2005AcA....55..123G} [67] Gazeas K.~D., Baran A., Niarchos P., Zola S., Kreiner J.~M., Ogloza W., Rucinski S.~M., et al., 2005, AcA, 55, 123
\bibitem[\protect\citeauthoryear{Yu et al.}{2017}]{2017NewA...55...13Y} [68] Yu Y.-X., Zhang X.-D., Hu K., Xiang F.-Y., 2017, NewA, 55, 13. doi:10.1016/j.newast.2017.02.002
\bibitem[\protect\citeauthoryear{Perryman}{1997}]{perryman1997hipparcos} [69] Perryman M., 1997, Noordwijk: ESA
\bibitem[\protect\citeauthoryear{Lu, Rucinski, \& Og{\l}oza}{2001}]{2001AJ....122..402L} [70] Lu W., Rucinski S.~M., Og{\l}oza W., 2001, AJ, 122, 402. doi:10.1086/321131
\bibitem[\protect\citeauthoryear{Pribulla et al.}{2002}]{2002IBVS.5258....1P} [71] Pribulla T., Chochol D., Vanko M., Parimucha S., 2002, IBVS, 5258, 1
\bibitem[\protect\citeauthoryear{D'Angelo, van Kerkwijk, \& Rucinski}{2006}]{2006AJ....132..650D} [72] D'Angelo C., van Kerkwijk M.~H., Rucinski S.~M., 2006, AJ, 132, 650. doi:10.1086/505265
\bibitem[\protect\citeauthoryear{Balanowsky}{1922}]{1922AN....216..159B} [73] Balanowsky I., 1922, AN, 216, 159. doi:10.1002/asna.19222161006
\bibitem[\protect\citeauthoryear{Rucinski, Lu, \& Mochnacki}{2000}]{2000AJ....120.1133R} [74] Rucinski S.~M., Lu W., Mochnacki S.~W., 2000, AJ, 120, 1133. doi:10.1086/301458
\bibitem[\protect\citeauthoryear{Zola et al.}{2005}]{2005AcA....55..389Z} [75] Zola S., Kreiner J.~M., Zakrzewski B., Kjurkchieva D.~P., Marchev D.~V., Baran A., Rucinski S.~M., et al., 2005, AcA, 55, 389
\bibitem[\protect\citeauthoryear{Kaitchuck et al.}{2006}]{2006JAVSO..34..165K} [76] Kaitchuck R.~H., Hill R.~L., Corn A.~P., Gevirtz J., Levell K.~L., Valenti T.~L., 2006, JAVSO, 34, 165
\bibitem[\protect\citeauthoryear{Liu et al.}{2012}]{2012PASJ...64...48L} [77] Liu L., Qian S.-B., He J.-J., Liao W.-P., Zhu L.-Y., Zhao E., 2012, PASJ, 64, 48. doi:10.1093/pasj/64.3.48
\bibitem[\protect\citeauthoryear{Hutton}{1992}]{1992IBVS.3723....1H} [78] Hutton R.~G., 1992, IBVS, 3723, 1
\bibitem[\protect\citeauthoryear{Schirmer \& Geyer}{1992}]{1992IBVS.3785....1S} [79] Schirmer J., Geyer E.~H., 1992, IBVS, 3785, 1
\bibitem[\protect\citeauthoryear{Robb}{1992}]{1992IBVS.3798....1R} [80] Robb R.~M., 1992, IBVS, 3798, 1
\bibitem[\protect\citeauthoryear{Molik \& Wolf}{2000}]{2000IBVS.4951....1M} [81] Molik P., Wolf M., 2000, IBVS, 4951, 1
\bibitem[\protect\citeauthoryear{Gazeas, Niarchos, \& Zola}{2007}]{2007ASPC..370..279G} [82] Gazeas K.~D., Niarchos P.~G., Zola S., 2007, ASPC, 370, 279
\bibitem[\protect\citeauthoryear{Li \& Qian}{2013}]{2013NewA...21...46L} [83] Li K., Qian S.-B., 2013, NewA, 21, 46. doi:10.1016/j.newast.2012.11.003
\bibitem[\protect\citeauthoryear{Weber}{1963}]{1963IBVS...21....1W} [84] Weber R., 1963, IBVS, 21, 1
\bibitem[\protect\citeauthoryear{Diethelm \& Gautschy}{1980}]{1980IBVS.1767....1D} [85] Diethelm R., Gautschy A., 1980, IBVS, 1767, 1
\bibitem[\protect\citeauthoryear{G{\"u}rol \& M{\"u}yessero{\u{g}}lu}{2005}]{2005AN....326...43G} [86] G{\"u}rol B., M{\"u}yessero{\u{g}}lu Z., 2005, AN, 326, 43. doi:10.1002/asna.200410338
\bibitem[\protect\citeauthoryear{Nelson \& Robb}{2015}]{2015IBVS.6134....1N} [87] Nelson R.~H., Robb R.~M., 2015, IBVS, 6134, 1
\bibitem[\protect\citeauthoryear{Huang et al.}{2021}]{2021RAA....21..120H} [88] Huang H.-P., Yu Y.-X., Yu T., Hu K., Xiang F.-Y., 2021, RAA, 21, 120. doi:10.1088/1674-4527/21/5/120
\bibitem[\protect\citeauthoryear{Pribulla \& Vanko}{2002}]{2002CoSka..32...79P} [89] Pribulla T., Vanko M., 2002, CoSka, 32, 79
\bibitem[\protect\citeauthoryear{Jordan}{1923}]{1923AJ.....35...44J} [90] Jordan F.~C., 1923, AJ, 35, 44. doi:10.1086/104568
\bibitem[\protect\citeauthoryear{Struve}{1950}]{1950ApJ...111Q.658S} [91] Struve O., 1950, ApJ, 111, 658. doi:10.1086/145313
\bibitem[\protect\citeauthoryear{Milone et al.}{1980}]{1980ApJS...43..339M} [92] Milone E.~F., Chia T.~T., Castle K.~G., Robb R.~M., Merrill J.~E., 1980, ApJS, 43, 339. doi:10.1086/190671
\bibitem[\protect\citeauthoryear{Milone et al.}{1985}]{1985AJ.....90..109M} [93] Milone E.~F., Hrivnak B.~J., Hill G., Fisher W.~A., 1985, AJ, 90, 109. doi:10.1086/113716
\bibitem[\protect\citeauthoryear{Srivastava}{1987}]{1987ApSS.139..373S} [94] Srivastava R.~K., 1987, Ap\&SS, 139, 373. doi:10.1007/BF00644366
\bibitem[\protect\citeauthoryear{Qian}{2002}]{2002AA...384..908Q} [95] Qian S., 2002, A\&A, 384, 908. doi:10.1051/0004-6361:20020108
\bibitem[\protect\citeauthoryear{Rucinski, Pribulla, \& van Kerkwijk}{2007}]{2007AJ....134.2353R} [96] Rucinski S.~M., Pribulla T., van Kerkwijk M.~H., 2007, AJ, 134, 2353. doi:10.1086/523353
\bibitem[\protect\citeauthoryear{Pribulla et al.}{2009}]{2009AJ....137.3646P} [97] Pribulla T., Rucinski S.~M., DeBond H., De Ridder A., Karmo T., Thomson J.~R., Croll B., et al., 2009, AJ, 137, 3646. doi:10.1088/0004-6256/137/3/3646
\bibitem[\protect\citeauthoryear{Djura{\v{s}}evi{\'c} et al.}{2011}]{2011AA...525A..66D} [98] Djura{\v{s}}evi{\'c} G., Y{\i}lmaz M., Ba{\c{s}}t{\"u}rk {\"O}., K{\i}l{\i}{\c{c}}o{\u{g}}lu T., Latkovi{\'c} O., {\c{C}}al{\i}{\c{s}}kan {\c{S}}., 2011, A\&A, 525, A66. doi:10.1051/0004-6361/201014895
\bibitem[\protect\citeauthoryear{Leavitt}{1908}]{1908AnHar..60...87L} [99] Leavitt H.~S., 1908, AnHar, 60, 87
\bibitem[\protect\citeauthoryear{Hertzsprung}{1928}]{1928BAN.....4..153H} [100] Hertzsprung E., 1928, BAN, 4, 153
\bibitem[\protect\citeauthoryear{Marton, Grieco, \& Sistero}{1989}]{1989MNRAS.240..931M} [101] Marton S.~F., Grieco A., Sistero R.~F., 1989, MNRAS, 240, 931. doi:10.1093/mnras/240.4.931
\bibitem[\protect\citeauthoryear{Kaluzny \& Caillault}{1989}]{1989AcA....39...27K} [102] Kaluzny J., Caillault J.-P., 1989, AcA, 39, 27
\bibitem[\protect\citeauthoryear{Hilditch, Hill, \& Bell}{1992}]{1992MNRAS.255..285H} [103] Hilditch R.~W., Hill G., Bell S.~A., 1992, MNRAS, 255, 285. doi:10.1093/mnras/255.2.285
\bibitem[\protect\citeauthoryear{Duerbeck \& Rucinski}{2007}]{2007AJ....133..169D} [104] Duerbeck H.~W., Rucinski S.~M., 2007, AJ, 133, 169. doi:10.1086/509764
\bibitem[\protect\citeauthoryear{Sarotsakulchai et al.}{2019}]{2019PASJ...71...34S} [105] Sarotsakulchai T., Qian S.-B., Soonthornthum B., Fern{\'a}ndez Laj{\'u}s E., Liu N.-P., Zhou X., Zhang J., et al., 2019, PASJ, 71, 34. doi:10.1093/pasj/psy149
\bibitem[\protect\citeauthoryear{Gaposchkin}{1932}]{1932VeBB....9D...1G} [106] Gaposchkin S., 1932, VeBB, 9, D1
\bibitem[\protect\citeauthoryear{Gaposchkin}{1938}]{payne1938variable} [107] Gaposchkin S., 1938, Harvard Monograph, v.5
\bibitem[\protect\citeauthoryear{Struve \& Gratton}{1948}]{1948ApJ...108..497S} [108] Struve O., Gratton L., 1948, ApJ, 108, 497. doi:10.1086/145086
\bibitem[\protect\citeauthoryear{Kopal}{1955}]{1955AnAp...18..379K} [109] Kopal Z., 1955, AnAp, 18, 379
\bibitem[\protect\citeauthoryear{Xiang \& Zhou}{2004}]{2004NewA....9..273X} [110] Xiang F.~Y., Zhou Y.~C., 2004, NewA, 9, 273. doi:10.1016/j.newast.2003.12.001
\bibitem[\protect\citeauthoryear{Nelson \& Alton}{2019}]{2019IBVS.6266....1N} [111] Nelson R.~H., Alton K.~B., 2019, IBVS, 6266, 1. doi:10.22444/IBVS.6266
\bibitem[\protect\citeauthoryear{Hoffmeister}{1933}]{1933AN....247..281H} [112] Hoffmeister C., 1933, AN, 247, 281. doi:10.1002/asna.19322471502
\bibitem[\protect\citeauthoryear{Zessewitsch}{1944}]{zessewitsch1944} [113] Zessewitsch W.P., 1944, Kazan Astron. Circ., 32, 6
\bibitem[\protect\citeauthoryear{Struve et al.}{1950}]{1950ApJ...111R.658S} [114] Struve O., Horak H.~G., Canavaggia R., Kourganoff V., Colacevich A., 1950, ApJ, 111, 658. doi:10.1086/145314
\bibitem[\protect\citeauthoryear{Cillie}{1951}]{1951AJ.....56...35C} [115] Cillie G.~G., 1951, AJ, 56, 35. doi:10.1086/106484
\bibitem[\protect\citeauthoryear{Archer}{1959}]{1959ApJ...130..774A} [116] Archer S., 1959, ApJ, 130, 774. doi:10.1086/146769
\bibitem[\protect\citeauthoryear{Russo et al.}{1982}]{1982AAS...47..211R} [117] Russo G., Sollazzo C., Maceroni C., Milano L., 1982, A\&AS, 47, 211
\bibitem[\protect\citeauthoryear{Strohmeier et al.}{1960}]{strohmeier1960veroff} [118] Strohmeier W., Knigge R., Ott H., 1960, V (5)
\bibitem[\protect\citeauthoryear{Bossen}{1973}]{1973AAS...10..217B} [119] Bossen H., 1973, A\&AS, 10, 217
\bibitem[\protect\citeauthoryear{Markworth \& Michaels}{1982}]{1982PASP...94..350M} [120] Markworth N.~L., Michaels E.~J., 1982, PASP, 94, 350. doi:10.1086/130989
\bibitem[\protect\citeauthoryear{Zhang et al.}{1995}]{1995IBVS.4240....1Z} [121] Zhang X., Zhang R., Zhai D., Fang M., 1995, IBVS, 4240, 1
\bibitem[\protect\citeauthoryear{Vanko et al.}{2001}]{2001CoSka..31..129V} [122] Vanko M., Pribulla T., Chochol D., Parimucha S., Kim C.~H., Lee J.~W., Han J.~Y., 2001, CoSka, 31, 129
\bibitem[\protect\citeauthoryear{Soloviev}{1937}]{soloviev1937tadjik} [123] Soloviev A., 1937, Circ
\bibitem[\protect\citeauthoryear{Binnendijk}{1967}]{1967AJ.....72...82B} [124] Binnendijk L., 1967, AJ, 72, 82. doi:10.1086/110204
\bibitem[\protect\citeauthoryear{Mauder}{1972}]{1972AA....17....1M} [125] Mauder H., 1972, A\&A, 17, 1
\bibitem[\protect\citeauthoryear{Qian et al.}{2007}]{2007AJ....134.1769Q} [126] Qian S.-B., Yuan J.-Z., Xiang F.-Y., Soonthornthum B., Zhu L.-Y., He J.-J., 2007, AJ, 134, 1769. doi:10.1086/521927
\bibitem[\protect\citeauthoryear{Hoffmeister}{1935}]{1935AN....255..401H} [127] Hoffmeister C., 1935, AN, 255, 401. doi:10.1002/asna.19352552202
\bibitem[\protect\citeauthoryear{Rainger, Bell, \& Hilditch}{1990}]{1990MNRAS.246...47R} [128] Rainger P.~P., Bell S.~A., Hilditch R.~W., 1990, MNRAS, 246, 47
\bibitem[\protect\citeauthoryear{Qian \& Zhu}{2002}]{2002ApJ...568.1004Q} [129] Qian S.~B., Zhu L.~Y., 2002, ApJ, 568, 1004. doi:10.1086/338964
\bibitem[\protect\citeauthoryear{Liu et al.}{2011}]{2011AJ....141..147L} [130] Liu L., Qian S.-B., Zhu L.-Y., He J.-J., Li L.-J., 2011, AJ, 141, 147. doi:10.1088/0004-6256/141/5/147
\bibitem[\protect\citeauthoryear{Pr{\v{s}}a et al.}{2016}]{2016ApJS..227...29P} [131] Pr{\v{s}}a A., Conroy K.~E., Horvat M., Pablo H., Kochoska A., Bloemen S., Giammarco J., et al., 2016, ApJS, 227, 29. doi:10.3847/1538-4365/227/2/29
\bibitem[\protect\citeauthoryear{Conroy et al.}{2020}]{2020ApJS..250...34C} [132] Conroy K.~E., Kochoska A., Hey D., Pablo H., Hambleton K.~M., Jones D., Giammarco J., et al., 2020, ApJS, 250, 34. doi:10.3847/1538-4365/abb4e2
\bibitem[\protect\citeauthoryear{Ruci{\'n}ski}{1969}]{1969AcA....19..245R} [133] Ruci{\'n}ski S.~M., 1969, AcA, 19, 245
\bibitem[\protect\citeauthoryear{Lucy}{1967}]{1967ZA.....65...89L} [134] Lucy L.~B., 1967, ZA, 65, 89
\bibitem[\protect\citeauthoryear{Castelli \& Kurucz}{2004}]{castelli2004new} [135] Castelli, Fiorella and Kurucz, Robert L., arXiv preprint astro-ph/0405087
\bibitem[\protect\citeauthoryear{Sriram et al.}{2017}]{2017AJ....153..231S} [136] Sriram K., Malu S., Choi C.~S., Vivekananda Rao P., 2017, AJ, 153, 231. doi:10.3847/1538-3881/aa6893
\bibitem[\protect\citeauthoryear{Tavakkoli et al.}{2015}]{2015NewA...37...64T} [137] Tavakkoli F., Hasanzadeh A., Poro A., 2015, NewA, 37, 64. doi:10.1016/j.newast.2014.12.004
\bibitem[\protect\citeauthoryear{Poro et al.}{2023}]{2023RAA....23i5011P} [138] Poro A., Fern{\'a}ndez-Laj{\'u}s E., Madani M., Sabbaghian G., Nasrollahzadeh F., Jahediparizi F., 2023, RAA, 23, 095011. doi:10.1088/1674-4527/ace027
\bibitem[\protect\citeauthoryear{Poro et al.}{2024}]{2024PASP..136b4201P} [139] Poro A., Tanriver M., Michel R., Paki E., 2024, PASP, 136, 024201. doi:10.1088/1538-3873/ad1ed3
\bibitem[\protect\citeauthoryear{Poro et al.}{2024}]{2024NewA..11002227P} [140] Poro A., Hedayatjoo M., Nastaran M., Nourmohammad M., Azarara H., et al., 2024, NewA, 110, 102227. doi:10.1016/j.newast.2024.102227
\bibitem[\protect\citeauthoryear{Poro et al.}{2021}]{2021NewA...8601571P} [141] Poro A., Zamanpour S., Hashemi M., Alada{\u{g}} Y., Aksaker N., Rezaei S., Solmaz A., 2021, NewA, 86, 101571. doi:10.1016/j.newast.2021.101571
\bibitem[\protect\citeauthoryear{Kouzuma}{2023}]{2023ApJ...958...84K} [142] Kouzuma S., 2023, ApJ, 958, 84. doi:10.3847/1538-4357/ad03e1
\bibitem[\protect\citeauthoryear{Torres}{2010}]{2010AJ....140.1158T} [143] Torres G., 2010, AJ, 140, 1158. doi:10.1088/0004-6256/140/5/1158
\bibitem[\protect\citeauthoryear{Cox}{2000}]{2000asqu.book.....C} [144] Cox A.~N., 2000, asqu.book
\bibitem[\protect\citeauthoryear{Gaia Collaboration et al.}{2023}]{2023AA...674A..33G} [145] Gaia Collaboration, Montegriffo P., Bellazzini M., De Angeli F., Andrae R., Barstow M.~A., et al., 2023, A\&A, 674, A33. doi:10.1051/0004-6361/202243709
\bibitem[\protect\citeauthoryear{Li et al.}{2021}]{2021ApJ...922..122L} [146] Li K., Xia Q.-Q., Kim C.-H., Hu S.-M., Guo D.-F., et al., 2021, ApJ, 922, 122. doi:10.3847/1538-4357/ac242f
\bibitem[\protect\citeauthoryear{Qian et al.}{2005}]{2005AJ....130..224Q} [147] Qian S.-B., Yang Y.-G., Soonthornthum B., Zhu L.-Y., He J.-J., Yuan J.-Z., 2005, AJ, 130, 224. doi:10.1086/430673
\bibitem[\protect\citeauthoryear{Yang et al.}{2009}]{2009AJ....138..540Y} [148] Yang Y.-G., Qian S.-B., Zhu L.-Y., He J.-J., 2009, AJ, 138, 540. doi:10.1088/0004-6256/138/2/540
\bibitem[\protect\citeauthoryear{Qian et al.}{2011}]{2011AJ....141..151Q} [149] Qian S.-B., Liu L., Zhu L.-Y., He J.-J., Yang Y.-G., Bernasconi L., 2011, AJ, 141, 151. doi:10.1088/0004-6256/141/5/151
\bibitem[\protect\citeauthoryear{Liao et al.}{2017}]{liao2017} [150] Liao, W-P and Qian, S-B and Soonthornthum, B and Sarotsakulchai, T and Zhu, L-Y and Zhang, J and Irina, Voloshina, 2017, PASP, 129, 124204
\bibitem[\protect\citeauthoryear{Eker et al.}{2006}]{2006MNRAS.373.1483E} [151] Eker Z., Demircan O., Bilir S., Karata{\c{s}} Y., 2006, MNRAS, 373, 1483. doi:10.1111/j.1365-2966.2006.11073.x
\bibitem[\protect\citeauthoryear{Poro et al.}{2024}]{2024RAA....24e5001P} [152] Poro A., Baudart S., Nourmohammad M., Sabaghpour Arani Z., Farhadi F., et al., 2024, RAA, 24, 055001. doi:10.1088/1674-4527/ad3a2c
\end{thebibliography}
\end{document}